\documentstyle[12pt,a4wide]{article} 

\newcommand{\news}{\setcounter{equation}{0}}

\def\eqn{\begin{equation}} 
\def\eeqn{\end{equation}}
\def\arr{\begin{array}} 
\def\earr{\end{array}}
\def\eqna{\begin{eqnarray}} 
\def\eeqna{\end{eqnarray}} 
\def\a{\alpha}
\def\b{\beta} 
\def\D{\Delta} 
\def\s{\sigma} 
\def\d{\delta}
 
\def\o{\omega} 
\def\O{\Omega} 
\def\e{\epsilon}
\def\th{\theta} 
 
\def\n{\nu} 
\def\la{\lambda} 
\def\z{\zeta}
 
\def\p{\partial} 
 
\def\ha{\hat{\alpha}}
\def\hb{\hat{\beta}} 
\def\hg{\hat{\gamma}} 
\def\hth{\hat{\theta}}
\font\mybb=msbm10 at 12pt 
\def\bb#1{\hbox{\mybb#1}}

\def\bE {\bb{E}}

\begin{document}

\vspace*{-.6in}
\thispagestyle{empty}
\begin{flushright}
DAMTP-1998-75\\
\end{flushright}

{\Large 
\begin{center}
{\bf Black hole dynamics from instanton strings}
\end{center}}
\vspace{.3in}
\begin{center}
Miguel S. Costa\footnote{M.S.Costa@damtp.cam.ac.uk}\\
\vspace{.1in}
\emph{D.A.M.T.P.\\ University of Cambridge \\ Cambridge CB3 9EW \\ UK}
\end{center}

\vspace{.5in}

\begin{abstract}
A D-5-brane bound state with a self-dual field strength on a 4-torus 
is considered. In a particular case this model reproduces the D5-D1 brane
bound state usually used in the string theory description of 5-dimensional 
black holes. In the limit where the brane dynamics decouples 
from the bulk the Higgs and Coulomb branches of the theory on the 
brane decouple. Contrasting with the usual instanton moduli space 
approximation to the problem the Higgs branch describes fundamental 
excitations of the gauge field on the brane. Upon reduction to 
2-dimensions it is associated with the so-called instanton strings. Using 
the Dirac-Born-Infeld action for the D-5-brane we determine the coupling
of these strings to a minimally coupled scalar in the black hole background.
The supergravity calculation of the cross section is found to agree with 
the D-brane absorption probability rate calculation. We consider the
near horizon geometry of our black hole and elaborate on the corresponding
duality with the Higgs branch of the gauge theory in the large $N$ limit.
A heuristic argument for the scaling of the effective string
tension is given.
\end{abstract}
\newpage

\section{Introduction}
\news
Over the past two years several issues in
black hole physics have been  successfully addressed within the
framework of string theory (see  \cite{Youm,Peet} for reviews and
complete lists of references). The  black hole dynamics may be
recovered from an effective string description \cite{StroVafa,Mald1,DasMath}. 
In the dilute gas approximation \cite{MaldStro}, i.e. when the left- and 
right-moving modes on the effective string are free 
and when anti-branes are suppressed, the Bekenstein-Hawking entropy 
is correctly reproduced. Further, assuming that this effective string 
couples to the bulk fields with a Dirac-Born-Infeld type action it has been 
possible to find agreement with the classical cross section calculations 
for scalar and fermionic bulk fields [5-18]. These calculations
provide a highly non-trivial test of the effective string
model. However, the derivation of the effective string action including its
coupling to the bulk fields requires several
assumptions. In other words, we would like to deduce this action from
first principles as it is the case for similar calculations involving
the D-3-brane \cite{Kleb,Kleb..,KlebGubs}. One of the purposes of 
this paper is to fill in this gap.

We shall  consider the D-5-brane bound state with a constant self-dual
worldvolume field strength on a compact $T^4$ studied in
\cite{CostaPerry}. This configuration includes as a special case the 
D5-D1 brane bound state used in the original
derivation of the Bekenstein-Hawking entropy formula \cite{StroVafa}.
The gauge theory fluctuating spectrum associated with this bound state
was found to agree with the spectrum derived from open strings ending
on the D-5-brane bound state \cite{Polc,CostaPerry,CostaPerry1}. For this
reason the modes associated with the worldvolume fields should be 
regarded as fundamental excitations of the D-brane system. This includes
some modes of the gauge field that are self-dual on $T^4$ and may be
called instantons but should not be interpreted as solitons. We shall
see that in the limit where the brane dynamics decouples from the bulk
we may define two supersymmetric branches of the theory on the brane 
corresponding to the self-dual modes and to the modes associated with
the movement of the brane system in the transverse directions. They
define the Higgs and Coulomb branches of the theory that are shown to
decouple in the above limit. The Higgs branch is the one associated
with the dynamics of the black hole. We derive from first principles
the action for the bosonic fields in the Higgs branch which we call
instanton strings action rather then effective string action.
We also consider the coupling of these instanton strings to a minimally
coupled scalar in the black hole background, finding agreement with the
scattering cross section calculation on the supergravity side. This 
agreement follows because both string and classical calculations have
an overlapping domain of validity (this will be our analogue of the
double scaling limit introduced by Klebanov \cite{Kleb}), giving a 
rationale for why both descriptions yield the same result. A deeper 
explanation is uncovered by Maldacena's duality proposal \cite{Mald2} 
and subsequent works \cite{Gubs..,Witt3}.
We shall elaborate on this proposal.
In particular, we argue that {\em the Higgs branch of the large $N$ 
limit of 6-dimensional super Yang-Mills theory with a 't Hooft
twist on a compact $T^4$ is dual to supergravity on 
$AdS_3\times S^3\times T^4$}. Based on this interpretation of the
duality conjecture the effective string action should be associated
with this large $N$ limit of the theory. We shall give a heuristic 
derivation of the effective string tension which agrees with 
previous results \cite{Math,Gubs,HassWadia}.

The paper is organised as follows: In section two we shall revise the
model studied in \cite{CostaPerry} and analyse the brane dynamics when 
it decouples from the bulk. The regions of validity of both D-brane and 
supergravity approximations are explained. In section three we shall 
find a minimally coupled scalar
in our black hole background and derive the corresponding coupling to
the instanton strings. Section four is devoted to the supergravity
calculation of the scattering cross section as well as the corresponding
D-brane absorption probability rate. In section five we shall describe 
the double scaling limit where both calculations are expected to agree. 
After analysing the near horizon geometry associated with our black
hole we consider Maldacena's duality proposal. We give our conclusions 
in section six.

\section{The model}
\news

In this section we shall review the D-brane model associated with our
five-dimensional black hole. The dynamics of the D-brane system will be
derived by starting from the super Yang-Mills (SYM) action. We shall 
comment on the validity of such approximation. We review the fluctuating
spectrum, study the decoupling of the Higgs and Coulomb branches of the
theory when the brane dynamics decouples from the bulk and derive the 
action for the instanton strings determining the
black hole dynamics. We then write the supergravity 
solution describing the geometry of our black hole and comment on the
validity of the supergravity approximation.

Because we are claiming that our model also describes the D5-D1 brane
bound state we shall keep referring to this special case as we proceed.

\subsection{D-brane phase}

We consider a bound state of two D-5-branes wrapped on $S^1\times T^4$
with coordinates $x^1,...,x^5$ (the generalisation to the case of $n$
D-5-branes is straightforward). Each D-5-brane has winding numbers $N_i$
along $S^1$, $p_i$ along the $x^2$-direction and $\bar{p}_i$ along the
$x^4$-direction. Thus, the worldvolume fields take values on the 
$U(N_1p_1\bar{p}_1+N_2p_2\bar{p}_2)$ Lie algebra \cite{Witt4}. 
In order to have a
non-trivial D-5-brane configuration we turn on the worldvolume gauge 
field such that the corresponding field strength is diagonal and
self-dual on $T^4$. The non-vanishing components are taken to be
(we assume without loss of generality that $\tan{\th_1}>\tan{\th_2}$)
\eqna
G^0_{23}=G^0_{45}=\frac{1}{2\pi\a'}
{\rm diag}\left(\underbrace{\tan{\th_1},...,
\tan{\th_1}},\underbrace{\tan{\th_2},...,\tan{\th_2}}\right)\ ,
\nonumber\\
{\footnotesize N_1p_1\bar{p}_1
\ {\rm times}\ \ \ \ \ \ N_2p_2\bar{p}_2\ {\rm times}\ \ \ \ \ \ \ \ }
\label{2.1}
\eeqna
where 
\eqn
\frac{1}{2\pi\a'}\tan{\th_i}=\frac{2\pi}{L_2L_3}\frac{q_i}{p_i}=
\frac{2\pi}{L_4L_5}\frac{\bar{q}_i}{\bar{p}_i}\ ,
\label{2.2}
\eeqn
with $q_i$ and $\bar{q}_i$ integers and $L_{\ha}=2\pi R_{\ha}$ the 
length of each $T^4$ circle ($\ha=2,...,5$). This vacuum expectation
value for the gauge field breaks the gauge invariance to 
$U(N_1p_1\bar{p}_1)\otimes U(N_2p_2\bar{p}_2)$. Because the branes are
wrapped along the $x^1$-, $x^2$- and $x^4$-directions the gauge invariance
is further broken to $U(1)^{N_1p_1\bar{p}_1+N_2p_2\bar{p}_2}$. Each
D-5-brane carries $Q_{5_i}=N_ip_i\bar{p}_i$ units of D-5-brane charge.
Thus, the total D-5-brane charge is
\eqn
Q_5=N_1p_1\bar{p}_1+N_2p_2\bar{p}_2\ .
\label{Q5}
\eeqn
Each brane carries fluxes in the $x^2x^3$ and $x^4x^5$ 2-tori. The total
fluxes are
\eqn
\arr{l}
{\cal F}_{23}={\displaystyle \frac{1}{2\pi}\int_{T^{^{2}}_{_{(23)}}}}
{\rm tr}\ G^0=
\left( N_1q_1\bar{p}_1+N_2q_2\bar{p}_2\right)\ ,\\
{\cal F}_{45}={\displaystyle \frac{1}{2\pi}\int_{T^{^{2}}_{_{(45)}}}}
{\rm tr}\ G^0=
\left( N_1p_1\bar{q}_1+N_2p_2\bar{q}_2\right)\ .
\earr
\label{fluxes}
\eeqn
These fluxes induce a 't Hooft twist on the fields [30-33], i.e. the 
worldvolume fields obey twisted boundary conditions on $T^4$. Also, due 
to this vacuum expectation value for the field strength the D-5-branes
carry other D-brane charges. There are $Q_3={\cal F}_{45}$
D-3-brane charge units associated with D-3-branes parallel to the 
$(123)$-directions, and $Q_{3'}={\cal F}_{23}$
D-3-brane charge units associated with D-3-branes parallel to the
$(145)$-directions. Furthermore, the instanton number
associated with the background field strength is non-zero. As a
consequence the bound state carries the D-string charge \cite{Doug}
\eqn
Q_1=N_{ins}=\frac{1}{16\pi^2}\int_{T^4}{\rm tr}\ (G^0\wedge G^0)=
N_1q_1\bar{q}_1+N_2q_2\bar{q}_2\ .
\label{Q1}
\eeqn
It is now clear how we can obtain a bound state with the same charges 
as de D5-D1 brane system. We just have to set the fluxes in (\ref{fluxes})
to zero and the charges $Q_5$ and $Q_1$ are given by (\ref{Q5}) and 
(\ref{Q1}), respectively. For example, if we set $q_1=\bar{q}_1=1$ and
$q_2=\bar{q}_2=-1$, then $N_1p_1=N_2p_2$, $N_1\bar{p}_1=N_2\bar{p}_2$
and the D-string charge is $Q_1=N_1+N_2$. 

Now we consider the region of validity of the D-brane description of our 
bound state. Throughout this paper we shall always assume that $g\ll 1$
so closed string effects beyond tree level are suppressed. Also, we
assume that the size of $T^4$ is small, i.e. 
$L_{\ha}\sim\sqrt{\a'}$. The effective coupling constant for D-brane
string perturbation theory is usually $gN$ for $N$ D-branes on top of each
other. However, the presence of a condensate on the D-brane worldvolume 
induces a factor $\sqrt{1+(2\pi\a'G^0)^2}$ in the effective coupling 
\cite{Abou..}.
Thus, in our case the effective string coupling reads
\eqn
g_{eff}=gN_ip_i\bar{p}_i
\sqrt{1+(2\pi\a'G^0_{23})^2}\sqrt{1+(2\pi\a'G^0_{45})^2}
\equiv\frac{r_i^2}{\a'}\ .
\label{2.3}
\eeqn
The length scales $r_i$ will enter the supergravity solution below and
we assume for simplicity $r_1\sim r_2$.
D-brane perturbation theory is valid for \cite{MaldStro}
\eqn
r_i\ll 1\ ,
\label{2.4}
\eeqn
where the $r_i$ are now written in string units. In this region open string
loop corrections may be neglected and the dynamics for the low lying
modes on the brane is determined by the Dirac-Born-Infeld (DBI) action. Our
tool to study this region of parameters is the ten-dimensional
SYM action reduced to six dimensions. The corresponding
bosonic action is 
\eqn
S_{YM}=
-\frac{1}{g_{YM}^2}\int d^6x\ {\rm tr}\left\{\frac{1}{4}(G_{\a\b})^2+
\frac{1}{2}(\p_{\a}\phi_m+i[B_{\a},\phi_m])^2-
\frac{1}{4}[\phi_m,\phi_n]^2\right\}\ ,
\label{2.5}
\eeqn
where $\a,\b=0,...,5$ and $m,n=6,...,9$. We are taking the fields to be 
hermitian matrices with the field strength
given by $G_{\a\b}=\p_{\a}B_{\b}-\p_{\b}B_{\a}+i[B_{\a},B_{\b}]$.
The Yang-Mills coupling constant is related to the D-5-brane tension 
$T_5$ by 
\eqn
g_{YM}^2=\frac{1}{(2\pi \a')^2T_5}=(2\pi)^3g\a'\ .
\label{coupling}
\eeqn
Note that in our conventions both $B_{\a}$ and $\phi_m$ have the dimension 
of length$^{-1}$. This action is the leading term in the $\a'$ expansion 
of the DBI action. In this approximation we have
\eqn
\left(2\pi\a' G_{\ha\hb}^0\right)^2
\ll1\ \ \Rightarrow\ \  |\tan{\th_i}|\ll1\ .
\label{2.6}
\eeqn
Physically this condition may be obtained from the requirement
\eqn
M_{5_i}\gg M_{3_i},\ M_{3'_i},\ M_{1_i}\ ,
\label{2.7}
\eeqn
where $M_{3_i}$ and $M_{3'_i}$ are the masses of the D-3-branes
dissolved in the D-5-brane with mass $M_{5_i}$ and similarly
for $M_{1_i}$. If this condition does not hold we expect the D-5-branes
to be bent or deformed \cite{Mald3}. Because we are assuming that 
$L_{\ha}\sim \sqrt{\a'}$ we see from eqn. (\ref{2.2}) that the condition
(\ref{2.6}) gives $p_i\gg |q_i|$ and 
$\bar{p}_i\gg |\bar{q}_i|$. We remark that in this limit there is perfect
agreement between the string and the SYM spectrum derived in 
\cite{CostaPerry}.

Next we review the fluctuating spectra of the SYM theory. The starting
point is to expand the action around the background (\ref{2.1}). The
result is
\eqna
S_{SYM}&=& -\frac{1}{4g_{YM}^2}\int d^6x 
\ {\rm tr}\left\{-2A^{\a}D^2A_{\a}
-4iA^{\hb}[G_{\hb\ha}^0,A^{\ha}]-2\phi^mD^2\phi_m\right.
\nonumber\\\nonumber\\
&&\ \ \ \ \ \ \ \ \ \ \ \ +2i(D_{\a}A_{\b}-D_{\b}A_{\a})[A^{\b},A^{\a}]
+4i\phi^mD_{\a}[\phi_m,A^{\a}]
\label{2.8}\\\nonumber\\
&&\ \ \ \ \ \ \ \ \ \ \ \ \left. -[A_{\a},A_{\b}]^2-2[A_{\a},\phi_m]^2
-[\phi_m,\phi_n]^2\right\}\ ,\nonumber
\eeqna
where we have done the following splitting of the gauge field
\eqna
&&B_{\a}=B_{\a}^0+A_{\a}\ ,\ \ \ \ G_{\a\b}=G^0_{\a\b}+F_{\a\b}\ ,
\nonumber\\\nonumber\\
&&G^0_{\a\b}=\p_{\a}B_{\b}^0-\p_{\b}B_{\a}^0+i[B_{\a}^0,B_{\b}^0]\ ,
\nonumber\\\label{2.9}\\
&&F_{\a\b}=D_{\a}A_{\b}-D_{\b}A_{\a}+i[A_{\a},A_{\b}]\ ,
\nonumber
\eeqna
with $D_{\a}\equiv \p_{\a}+i[B_{\a}^0,\ \ ]$. The quantum fields $A_{\a}$
and $\phi_m$ are in the adjoint representation of 
$U(N_1p_1\bar{p}_1+N_2p_2\bar{p}_2)$ and $A_{\a}$ satisfies the background 
gauge fixing condition $D_{\a}A^{\a}=0$. These fields obey twisted
boundary conditions on $S^1\times T^4$ [30-33]. We have $(\b\ne 0)$
\eqn
A_{\a}(x^{\b}+L_{\b})=\O_{\b}A_{\a}(x^{\b})\O_{\b}^{-1}\ ,
\label{2.10}
\eeqn
and similarly for $\phi_m$. The $\O$'s are called multiple transition 
functions and take values on $U(N_1p_1\bar{p}_1+N_2p_2\bar{p}_2)$. They 
are given by 
\eqn
\O_{\a}={\rm Diag}(\O^{(1)}_{\a},\O^{(2)}_{\a})\ ,
\eeqn
where $\O^{(i)}_{\a}\in\ U(N_ip_i\bar{p}_i)$ and in terms of 
$U(p_i)\otimes U(\bar{p}_i)\otimes U(N_i)$ matrices reads
\eqna
&&\O_1^{(i)}={\bf 1}_{p_i}\otimes {\bf 1}_{\bar{p}_i}
\otimes V_{N_i}\ ,
\nonumber\\\nonumber\\
&&\O_2^{(i)}=\exp{\left[-\pi i n^i_{2\hb}x^{\hb}/L_{\hb}\right]}
{\bf V}_{p_i}\otimes {\bf 1}_{\bar{p}_i}\otimes {\bf 1}_{N_i}\ ,
\nonumber\\\nonumber\\
&&\O_3^{(i)}=\exp{\left[-\pi i n^i_{3\hb}x^{\hb}/L_{\hb}\right]}
({\bf U}_{p_i})^{q_i}\otimes {\bf 1}_{\bar{p}_i}\otimes {\bf 1}_{N_i}\ ,
\\\nonumber\\
&&\O_4^{(i)}=\exp{\left[-\pi i n^i_{4\hb}x^{\hb}/L_{\hb}\right]}
{\bf 1}_{p_i}\otimes {\bf V}_{\bar{p}_i}\otimes {\bf 1}_{N_i}\ ,
\nonumber\\\nonumber\\
&&\O_5^{(i)}=\exp{\left[-\pi i n^i_{5\hb}x^{\hb}/L_{\hb}\right]}
{\bf 1}_{p_i}\otimes ({\bf U}_{\bar{p}_i})^{\bar{q}_i}\otimes{\bf 1}_{N_i}\ .
\nonumber
\eeqna
The matrices ${\bf V}_{N_i}$ are the $N_i\times N_i$ shift matrices and
similarly for ${\bf V}_{p_i}$ and ${\bf V}_{\bar{p}_i}$. The matrices 
${\bf U}_{p_i}$ are given by
\eqn
{\bf U}_{p_i}={\rm diag}
\left( 1,e^{2\pi i\frac{1}{p_i}},...,e^{2\pi i\frac{p_i-1}{p_i}}\right)\ ,
\eeqn
and similarly for ${\bf U}_{\bar{p}_i}$. The tensors $n^i_{\ha\hb}$ are
called the twist tensors and are given by
\eqn
\arr{rl}
n^i_{\ha\hb}=&\left(
\arr{cccc}
0&q_i/p_i&0&0\\
-q_i/p_i&0&0&0\\
0&0&0&\bar{q}_i/\bar{p}_i\\
0&0&-\bar{q}_i/\bar{p}_i&0
\earr
\right)\ .
\earr
\eeqn
In order to analyse the spectrum it is convenient to decompose the fields
$A_{\a}$ and $\phi_m$ as
\eqn
A_{\a}=\left(
\arr{cc}
a_{\a}^1&b_{\a}\\
b_{\a}^{\dagger}&a_{\a}^2
\earr
\right)\ ,
\ \ \ \ \ \ \ \ 
\phi_m=\left(
\arr{cc}
c_m^1&d_m\\
d_m^{\dagger}&c_m^2
\earr
\right)\ .
\label{2.11}
\eeqn
The fields $a^i_{\a}$ and $c^i_m$ are in the adjoint representation of
$U(N_ip_i\bar{p}_i)$ and the fields $b_{\a}$ and $d_m$ in the
fundamental representation of 
$U(N_1p_1\bar{p}_1)\otimes U(\overline{N_2p_2\bar{p}_2})$.
Substituting the ansatz (\ref{2.11}) in the action (\ref{2.8}) and 
keeping only the quadratic terms in the fields together with the 
boundary conditions (\ref{2.10}) we may derive the spectrum of the theory.
The result is resumed in table 1. Note that we are using the complex 
coordinates $z_k=(x^{2k}-ix^{2k+1})/\sqrt{2}$ with $k=1,2$ to express
the fields $b_{\ha}$. It is important to realize that the functions 
$\chi^r_{m_1m_2}$ determining the mode expansion of the fields $b_{\a}$
and $d_m$ on $T^4$ have the same ``status'' as the usual modes 
$e^{ik_{\ha}x^{\ha}}$. They form a basis for functions satisfying the
twisted boundary conditions obeyed by these fields on $T^4$ and are
expressed in terms of $\Theta$-functions. Also the quadratic operator
$\hat{M}$ is the analogue of $(\p_{\ha})^2$. Each eigenvalue of this
operator has a degeneracy $n_L\bar{n}_L$ associated with the number of
Landau levels in the system.

\begin{table}
\begin{center}
\begin{tabular}{|c|c|c|c|c|} \hline
{\small Fields} & {\small Quadr. operators} & {\small Modes} &
{\small On-shell cond.} & {\small No. of d.o.f.} \\ \hline \hline
$a_{\ha}^i$ & $-(\p_{\s})^2-(\p_{\ha})^2$ & 
$e^{ik_{\s}x^{\s}+ik_{\ha}x^{\ha}}$ & $k_{\s}^2+k_{\ha}^2=0$ &
$4N_ip_i\bar{p}_i$ \\ \hline
$b_{z_k}$ & $-(\p_{\s})^2+(\hat{M}-4\pi f)$ & 
$e^{ik_{\s}x^{\s}}\chi^r_{m_1m_2}(x^{\ha})$ & $k_{\s}^2+\la^-_{m_1m_2}=0$ &
$4n_L\bar{n}_L$ \\ \hline
$b_{\bar{z}_k}$ & $-(\p_{\s})^2+(\hat{M}+4\pi f)$ & 
$e^{ik_{\s}x^{\s}}\chi^r_{m_1m_2}(x^{\ha})$ & $k_{\s}^2+\la^+_{m_1m_2}=0$ &
$4n_L\bar{n}_L$ \\ \hline
$c_m^i$ & $-(\p_{\s})^2-(\p_{\ha})^2$ & 
$e^{ik_{\s}x^{\s}+ik_{\ha}x^{\ha}}$ & $k_{\s}^2+k_{\ha}^2=0$ &
$4N_ip_i\bar{p}_i$ \\ \hline
$d_m$ & $-(\p_{\s})^2+\hat{M}$ & 
$e^{ik_{\s}x^{\s}}\chi^r_{m_1m_2}(x^{\ha})$ & $k_{\s}^2+\la_{m_1m_2}=0$ &
$8n_L\bar{n}_L$ \\ \hline
\end{tabular}
\end{center}
\caption{{\small Spectrum of the theory presented in a form suitable for
reduction to two dimensions. We have imposed the Coulomb gauge condition
$A_0=0$ and used the fact that $D_{\a}A^{\a}=0$ to fix $A_1$ (for the 
mode of $b_{z_k}$ with $\la^-=0$ this gives $A_1=0$).
The operator $\hat{M}$ is given by 
$\hat{M}=(i\p_{\ha}+\pi J_{\ha\hb}x^{\hb})^2$ with
$J_{\ha\hb}=(n^1_{\ha\hb}-n^2_{\ha\hb})/(L_{\ha}L_{\hb})$. The functions
$\chi^r_{m_1m_2}$ are eigenfunctions of the operator $\hat{M}$ with
eigenvalues $\la_{m_1m_2}=4\pi f(m_1+m_2+1)$ where 
$4\pi f=(\tan{\th_1}-\tan{\th_2})/(\pi\a')$ and 
$\la^{\pm}_{m_1m_2}=\la_{m_1m_2}\pm 4\pi f$. The index $r$ 
in the functions $\chi^r_{m_1m_2}$ runs from
$1$ to $n_L\bar{n}_L$ with $n_L=|p_1q_2-p_2q_1|$ and 
$\bar{n}_L=|\bar{p}_1\bar{q}_2-\bar{p}_2\bar{q}_1|$. 
The index $\s$ runs from $0$ to $1$.}}
\end{table}

In table 1 we wrote the mode expansion for the various fields but some
care is necessary because each field carries Lie algebra indices. 
Consider first the case of the fields $b_{\ha}$ and $d_m$. 
The corresponding modes in the table are defined on a 
$S^1_{eff}\times T^4_{eff}$ 5-torus while a given $a\bar{b}$ 
Lie algebra component of these fields takes values on
$S^1\times T^4$ determined by a given segment of 
$S^1_{eff}\times T^4_{eff}$. The different Lie algebra components are then
related by the boundary conditions. In the case of these fields we have
$S^1_{eff}$ with a length $L_{eff}=N_1N_2L_1$ and $T^4_{eff}$ with radi
($p_1p_2R_2,R_3,\bar{p}_1\bar{p}_2R_4,R_5$). A similar comment applies
to the $a^i_{\ha}$ and $c^i_m$ fields but now $S^1_{eff}$ has radius
$N_iR_1$ and $T^4_{eff}$ has radi ($p_iR_2,R_3,\bar{p}_iR_4,R_5$).
To be more explicite we consider the modes on $T^4$ of the
fields $b_{z_k}$ with the lowest eigenvalue $\la^-=0$ (i.e. $m_1=m_2=0$).
These are the only modes coming from the fields $b_{\ha}$ and $d_m$ that
are associated with massless particles in two dimensions. We write a 
given Lie algebra component of the fields, say the $1\bar{1}$
component, as $(b_{\bar{z}_k}^{1\bar{1}}=0)$
\eqna
&&b_{z_1}^{1\bar{1}}=\frac{1}{2\pi\a'}\sum_{r=1}^{n_L\bar{n}_L}
\xi^r_1(x^{\s})\cdot\chi^r(x^{\ha})\ ,
\nonumber
\\\label{2.12}\\
&&b_{z_2}^{1\bar{1}}=\frac{1}{2\pi\a'}\sum_{r=1}^{n_L\bar{n}_L}
\xi^r_2(x^{\s})\cdot\chi^r(x^{\ha})\ .
\nonumber
\eeqna
The complex fields $\xi^r_k$ are defined 
on an effective circle with radius $R_{eff}=N_1N_2R_1$ and 
$\chi^r(x^{\ha})$ takes values on the $T^4_{eff}$ defined above. The 
fields $b_{z_k}^{1\bar{1}}$ take values on $S^1\times T^4$ and all the
other Lie algebra components may be obtained from this one by using
the boundary conditions (\ref{2.10}).\footnote{We are assuming here that
$p_1, p_2$ and $\bar{p}_1, \bar{p}_2$ are co-prime. It is not difficult
to drop this condition \cite{CostaPerry}.} It is important to realize that
the fields $b_{z_k}^{a\bar{b}}$ are operators in the quantum theory
and therefore the fields $\xi^r_k$ are also quantum operators.

\subsubsection{Theory on the brane in the decoupling limit}

We consider the limit where the brane dynamics decouples from the bulk. 
This limit corresponds to take $\a'\rightarrow 0$ and 
$N_i\rightarrow\infty$, i.e. we are considering the 
large $N$ theory in the infrared limit. Noting that the radi of $T^4$ scale 
as $\sqrt{\a'}$ we conclude that the massive Ka\l u\.{z}a-Klein modes on 
$T^4$ associated with the fields $a^i_{\ha}$ and $c_m^i$ decouple from the 
theory in the above limit. Also, with the exception of the massless modes
associated with the fields $b_{z_k}$ in (\ref{2.12}) all the excitations
associated with $b_{\ha}$ and $d_m$ decouple. Thus, we are left with 
the massless excitations associated with the fields $c_m^i(x^{\s})$, 
$a_{\ha}^i(x^{\s})$ and $\xi^r_k(x^{\s})$.
 
To analyse the resulting theory it is convenient to consider the T-dual
six-dimensional theory with worldvolume given by the string
directions $x^{\s}$ and by the tranverse space to the D-5-brane system 
$\bE^4$. We follow very closely the analysis given in \cite{Mald3}. 
In fact, our model provides an explicite realization of the results 
there derived. We start with $N=2$ SUSY in $D=6$ but the self-dual 
background field strength on the $T^4$ breaks half of the supersymmetries 
leaving $N=1$ SUSY in $D=6$. There are two possible multiplets, the vector 
multiplet and the hypermultiplet \cite{Mald5}. The fields $c_m^i$ 
correspond to the gauge independent degrees of freedom of gauge 
bosons and therefore fall into a vector multiplet. The fields 
$a_{\ha}^i$ are $4$ scalars and fall into a hypermultiplet. 
The resulting theory is just two copies of $10D$ SYM compactified 
on $T^4$, each copy with gauge group $U(N_ip_i\bar{p}_i)$. So far we have 
the field content of $N=2$ SUSY in $D=6$. The fields $c_m^i$ in the vector 
multiplets are left invariant under the $SO(4)_I$ rotational symmetry 
associated with the $T^4$ while the fields $a_{\ha}^i$ in the hypermultiplets 
transform as $\bf{\bar{4}}$ under this symmetry. The  theory is not 
$SO(4)_I\simeq SU(2)_L\otimes SU(2)_R$ invariant because the background
field strength breaks this symmetry. If this field was not (anti) self-dual
we would be left with a $U(1)\otimes U(1)$ symmetry corresponding to 
rotations in the $x^2, x^3$ and $x^4, x^5$ directions. For (anti) self-dual
fields this symmetry gets enhanced to $U(2)\simeq SU(2)\otimes U(1)$ 
\cite{Baal}. The action of this group in the $z_k, \bar{z}_k$ 
coordinates is generated by 
\eqn
\rm{i}\bf{\s_3}\otimes \bf{1},\ \ \rm{i}\bf{\s_3}\otimes \bf{\s_3},\ \ 
\rm{i}\bf{1}\otimes \bf{\s_2},\ \ \rm{i}\bf{\s_3}\otimes \bf{\s_1},\ \ 
\eeqn
where the first generator corresponds to the $U(1)$ factor and the 
$\sigma$'s are the Pauli matrices. Thus, the 
resulting $N=1$ theory as a $U(2)$ $R$-symmetry. We still have to consider 
the complex fields $\xi^r_k$. For each $r$, they describe $4$ scalar fields
and therefore fall into a hypermultiplet \cite{Dijk..}. The fields
$\xi^r=(\xi^r_1\ \xi^r_2)$ transform as $\bf{\bar{2}}$ under the
$U(2)$ $R$-symmetry.

The reduction of the theory to two dimensions results in a theory with
$N=4$ SUSY in $2D$. Now both the hypers and the vectors have $4$ scalar
fields. They are distinguished by the different transformation
properties under $R$-symmetries. The theory has an extra 
$SO(4)_E\simeq SU(2)_{\tilde{L}}\otimes SU(2)_{\tilde{R}}$ $R$-symmetry
that leaves the scalars in the hypermultiplets unchanged but acts on the 
scalars in the vector multiplets. This theory has two supersymmetric 
branches, the Higgs branch where the hypers are excited and the Coulomb 
branch where the vectors are excited. Supersymmetry implies
that there is no coupling between vector and hyper multiplets 
\cite{Mald3}. We shall describe the different branches of the theory 
in the next subsection, for now let us just note that in the Higgs phase
the fields $a_{\ha}^i$ condense and the only independent degrees of freedom
are associated with the fields $\xi^r_k$.

All this resembles the moduli space approximation to the dynamics
of the D5-D1 brane bound state. 
The Higgs branch describing the moduli
of instantons on $T^4$ and the Coulomb branch the fluctuations of the
system in the transverse space. However, there is a crucial difference
in our description. The modes of the quantum fields $b_{z_k}$ that 
survived the decoupling limit are self-dual on $T^4$ and in that
sense deserve to be called instantons but rather then being interpreted
as solitons they should be interpreted as fundamental modes of the
fields (just like the standard $e^{ik_{\ha}x^{\ha}}$ modes). In other 
words, we are {\em not} quantising the collective coordinates of a 
soliton (instanton). There are two reasons for this: Firstly, 
they are the field theory realization of the low lying modes 
corresponding to open strings with ends on the D-5-branes with 
a different background field strength. Thus, they are as 
fundamental as the other modes corresponding
to the $a^i_{\a}$ and $c^i_m$ fields associated with open strings
ending on the same D-5-branes. Secondly, these instantons do not really have
a size in the sense that there is no moduli associated with its size.
In fact, all the dependence of $b_{z_k}$ on $T^4$ is through 
$x^{\ha}/L_{\ha}$. Thus, if the volume of $T^4$ is scaled the fields 
scale uniformly \cite{Doug..}. Also, this means that  we can take the limit 
$L_{\ha}\rightarrow 0$ and the field configuration remains well defined.

There is a potential problem when we take the size of $T^4$ to be of
order one in string units. Because the fields $b_{z_k}$ are $x^{\ha}$
dependent we could expect that string derivative corrections to the
DBI (or SYM) action become important \cite{Kita,AndrTsey}. It turns out
that for $\Delta\th\equiv\th_1-\th_2\ll 1$ which holds when the DBI 
corrections are suppressed the derivative corrections are also 
suppressed. To see this recall that the wave functions 
$\chi_{m_1m_2}$ in table 1 may be generated by the creation 
operators $a_k^{\dagger}$ with \cite{CostaPerry}
\eqna
&&a_k^{\dagger}=\frac{1}{i\sqrt{2\pi f}}
\left(\p_{z_k}-\pi f\bar{z}_k\right)\ ,\ \ \ k=1,2,
\nonumber\\\label{operators}\\
&&a_l=\frac{1}{i\sqrt{2\pi f}}
\left(\p_{\bar{z}_l}+\pi fz_l\right)\ ,\ \ \ l=1,2,
\nonumber
\eeqna
and $[a_l,a_k^{\dagger}]=\d_{lk}$. We then have 
$\langle z,\bar{z}|m_1m_2\rangle=\chi_{m_1m_2}(z,\bar{z})$, where
$|m_1m_2\rangle$ is normalised to unit. Considering
for example a typical derivative correction term like
$\sqrt{\a'}\p_2\chi_0$ we obtain from (\ref{operators})
\eqn
\sqrt{\a'}\p_2\chi_0=\rm{i}\sqrt{\frac{\Delta\th}{4\pi}}
\left( \chi_{1,0}+
\sqrt{\frac{\Delta\th}{4\pi}}\frac{x^3}{\sqrt{\a'}}\chi_0\right)\ ,
\eeqn
which is negligible for $\Delta\th\ll 1$ 
(note that $4\pi f=\D\th/(\pi\a')$). Thus, our field theory 
description is valid for $V_4\sim\a'^2$. As an aside note that the 
disagreement between the string and gauge theory normalisation for the
masses of the excitations on the brane bound state may be due to this
derivative corrections. A fact that as been suggested in \cite{HashTayl}.

To summarise, rather then doing a moduli space approximation
we have a vacuum state defined by the background field strength 
$G^0_{\ha\hb}$ which is a (constant) instanton on $T^4$. 
The quantum fluctuations
around this vacuum are well defined by open strings ending on the
D-5-brane bound state and their low energy field theory realization
is resumed in table 1. We have a quantum mechanical description of the
excitations around the instanton vacuum state pretty much as the 
description of the D-5-brane/D-string configuration given by 
Callan and Maldacena \cite{CallMald}. In the decoupling limit 
$\a'\rightarrow 0$ we ended up with the quantum fields $c_m^i(x^{\s})$, 
$a_{\ha}^i(x^{\s})$ and $\xi^r_k(x^{\s})$.

\subsubsection{Higgs and Coulomb branches}

Now we describe the Higgs and Coulomb branches of the theory. We shall
see that in the decoupling limit here considered these branches
decouple \cite{Ahar,Witt}. We want to define a supersymmetric branch
of the theory on the brane such that it will describe the dynamics of the
black holes considered in the next section. These black holes will be 
appropriately identified with some state of the theory on the
brane which may or may not preserve some supersymmetry.

Since the fields $b_{z_k}^{a\bar{b}}$ originate a self-dual field 
strength on $T^4$ the resulting
compactified theory is supersymmetric. However, when we consider the 
interactions between these fields it is seen that the fluctuating 
field strength $F_{\ha\hb}$ is no longer self-dual. To next
order in the fields $b_{z_k}^{a\bar{b}}$ the self-duality condition 
holds if the fields $a^i_{\ha}$ condense. They are determined by
\eqna
&a^i_{\hb}={\Box}^{-1}\p_{\ha}S^i_{\ha\hb}\ ,
\nonumber\\\nonumber\\
&(S^1_{\ha\hb})^{ab}=-i\left[
\left( b_{\ha}^{a\bar{c}}b_{\hb}^{\dagger\bar{c}b}-
b_{\hb}^{a\bar{c}}b_{\ha}^{\dagger\bar{c}b}\right)
-\frac{1}{2}\e_{\ha\hb\hg\hth}
\left( b_{\hg}^{a\bar{c}}b_{\hth}^{\dagger\bar{c}b}-
b_{\hth}^{a\bar{c}}b_{\hg}^{\dagger\bar{c}b}\right)\right]\ ,
\label{2.13}\\\nonumber\\
&(S^2_{\ha\hb})^{ab}=-i\left[
\left( b_{\ha}^{\dagger\bar{a}c}b_{\hb}^{c\bar{b}}-
b_{\hb}^{\dagger\bar{a}c}b_{\ha}^{c\bar{b}}\right)
-\frac{1}{2}\e_{\ha\hb\hg\hth}
\left( b_{\hg}^{\dagger\bar{a}c}b_{\hth}^{c\bar{b}}-
b_{\hth}^{\dagger\bar{a}c}b_{\hg}^{c\bar{b}}\right)\right]\ ,
\nonumber
\eeqna
where ${\Box}\equiv \p^2_{\ha}$.
We could have $N_ip_i\bar{p}_i$ commuting components of the free fields
$a^i_{\hb}$ and still have self-duality. The boundary conditions 
(\ref{2.10}) imply that these components would have to be on the diagonal
or on a shifted diagonal when the fields are expressed in terms of 
$U(p_i)\otimes U(\bar{p}_i)\otimes U(N_i)$ matrices. This would
give $4$ massless particles defined on an effective length $N_iL_1$.
This contribution is subleading in the large $N$ limit considered here. The
condensation of the fields is a familiar fact. It corresponds to 
require the D-term $[A_{\ha},A_{\hb}]^2$ in the action (\ref{2.8}) 
to vanish which does not happen 
when we consider just the fields $b_{z_k}^{a\bar{b}}$ (the cubic 
term in $A_{\ha}$ vanishes). Further, when the fields $\xi^r_k$ are excited
the commutator term $[A_{\ha},\phi_m]^2$ in the action gives a mass
term to the $c^i_m$ fields and vice-versa. Thus, in the low energy limit
we have the Higgs branch with the fields $\xi^r_k$ excited and the
Coulomb branch with the fields $c^i_m$ excited.

A more careful analysis is as follows: We start by considering a classical
field configuration that defines a supersymmetric branch of the theory,
i.e. we consider the moduli space of supersymmetric classical vacua. 
This corresponds to set all the D-terms of the theory to zero. 
The D-terms are
\eqn
V_1=-{\rm tr}[\phi_m,\phi_n]^2\ ,\ \ \ 
V_2=-{\rm tr}[A_{\ha},\phi_m]^2\ ,\ \ \ 
V_3=-{\rm tr}[A_{\ha},A_{\hb}]^2\ .
\nonumber
\eeqn
The unusual minus signs are because we took our fields to be hermitian.
$V_3$ vanishes because the $a^i_{\ha}$ fields condense, therefore 
we are just left with $V_1$ and $V_2$. They become
\eqna
&V_1=-{\rm tr}[c^1_m,c^1_n]^2-{\rm tr}[c^2_m,c^2_n]^2\ ,
\nonumber\\\\
&V_2=-{\rm tr}[a^1_{\ha},c^1_m]^2-{\rm tr}[a^2_{\ha},c^2_m]^2
+2{\rm tr}[(c^1_m)^2b_{\ha}b^{\dagger}_{\ha}+
(c^2_m)^2b^{\dagger}_{\ha}b_{\ha}-2c^2_mb^{\dagger}_{\ha}c^1_mb_{\ha}]\ .
\nonumber
\eeqna
Now there are only two possibilities (apart from the trivial case
$\phi_m\sim {\bf 1}$): (1) $c^i_m=0$ and then $\xi^r_k$ may be generic.
This is the Higgs branch. (2) The $c^i_m$ are generic but all commute. 
Because the branes are wrapped the boundary conditions require that
these fields take the form
\eqn
c^i_m\sim(V_{p_i})^r\otimes (V_{\bar{p}_i})^s
\otimes (V_{N_i})^t\ ,
\label{diag}
\eeqn
where the $V$'s are the shift matrices and $r,s,t$ are integers.
The claim is that in order to vanish
$V_2$ we need $\xi^r_k=0$. This may be seen by noting that if 
$\xi^r_k\ne 0$ the condensate  formed by the fields $a^i_{\ha}$ will
give a non-vanishing ${\rm tr}[a^i_{\ha},c^i_m]^2$.

We conclude that classically we either have a Higgs or a Coulomb
branch. Quantum mechanically we consider fluctuations of the fields
around the classical vacua
that obey the D-flatness conditions. Each branch defines a different
superconformal field theory. This has to be the case because a $(4,4)$
superconformal field theory has a $SU(2)\otimes SU(2)$ group of
left- and right-moving symmetries that must leave the scalars in the
theory invariant \cite{Witt1}. In the Higgs branch this group
originates from the $SO(4)_E$ symmetry while in the Coulomb branch
from the $SO(4)_I$ symmetry (which is broken to $U(2)$ in the 
Higgs branch).

\subsubsection{Instanton strings action}

The action for the fields $\xi^r_k$ in the Higgs branch may be obtained
by replacing the field configurations corresponding to (\ref{2.12}) 
and (\ref{2.13}) in the action (\ref{2.8}). We normalise the 
functions $\chi^r$ according to
\eqn
\int_{T_{eff}^4} d^4x\left(\chi^r\right)^*\chi^s=
(2\pi\sqrt{\a'})^4\delta^{rs}\ ,\ \ \ \ \ r,s=1,...,n_L\bar{n}_L\ ,
\label{norm}
\eeqn
which is well defined in the limit $\a'\rightarrow 0$ because 
$R_{\ha}\sim\sqrt{\a'}$. Defining $4n_L\bar{n}_L$ real fields $\z^r$
from the $\xi^r_k$ complex fields and replacing the field configuration
corresponding to (\ref{2.12}) in the action (\ref{2.8}) we obtain after 
some algebra the following $1+1$-dimensional free action
\eqn
S=-\frac{T_{ins}}{2}\int dt\int_0^{L_{eff}}dx^1
\sum_{r=1}^{4n_L\bar{n}_L}\p_{\s}\z^r\p^{\s}\z^r\ ,
\label{action}
\eeqn
where
\eqn
T_{ins}=\frac{1}{2\pi\a'g}\ ,\ \ \ \ L_{eff}=2\pi R_1N_1N_2\ ,
\ \ \ \ f=4n_L\bar{n}_L\ ,
\label{tension}
\eeqn
are the instanton strings tension, the effective length and the number of
bosonic (and fermionic) species in our model, respectively. In order to 
compare these results with the effective string model for the D5-D1 
system used in the literature let us write the corresponding quantities
\eqn
T_{eff}=\frac{1}{2\pi\a'g\sqrt{Q_5Q_1}}\ ,\ \ \ \ L_{eff}=2\pi R_1Q_5Q_1\ ,
\ \ \ \ f=4\ .
\label{tension1}
\eeqn
The particular combination of $Q_1$ and $Q_5$ in $T_{eff}$ has been derived
in \cite{Math,Gubs,HassWadia}. We shall argue in section 5 that 
by taking the large $N$ limit of our field theory the instanton 
strings tension gets normalised reproducing an effective string tension 
which agrees with this prediction. Using the result 
$Q_5Q_1=N_1N_2n_L\bar{n}_L$ which holds for the D5-D1 system we see that our
results for $L_{eff}$ and $f$ are not necessarily in contradiction with 
(\ref{tension1}). Note that for the D5-D1 brane bound state
described by our field configuration we always have $f\ge 16$.\footnote{We 
remark that this fact may be related to the fact that our field strength 
background does not have a minimal integer instanton number. $N_{ins}$ is 
always a multiple of $2$. Generalising our results to arbitrary integer 
instanton number may allow the possibility $f=4$.} The case $f=16$ 
corresponds to the example given after equation (\ref{Q1}) if we set 
$p_i=\bar{p}_i=1$. If this is the case one could argue that our
results are not reliable because the DBI corrections are important. This
is certainly true but things may not be as bad as they look. The reason
is that the supersymmetric configuration that we have found depends 
exclusively on the boundary conditions satisfied by the fields and on
the self-duality condition. The former depends only on the gauge 
invariance of the theory and it is certainly independent of the specific
Lagrangian describing the dynamics of the system. The latter is 
sufficient to show that our field configuration preserves a fraction 
of the supersymmetries and it is also independent of the specific 
Lagrangian. In fact, the self-duality condition was shown in 
\cite{Brec} to be a sufficient condition to minimise the non-abelian 
DBI action proposed by Tseytlin \cite{Tsey}. These arguments together 
with the string analysis given in \cite{CostaPerry} provide evidence for the 
validity of the supersymmetric field configuration even when the DBI 
corrections are expected to be important. Thus, we do not expect 
$L_{eff}$ and $f$ to be altered. What changes is the interacting theory 
and not the free action (\ref{action}).

We should now worry about the supersymmetry completion of the action
(\ref{action}). In \cite{Mald3} it was shown that this action takes
the form 
\eqn
S=-\frac{T_{ins}}{2}\int d^2x 
\left(\ G_{rs}(\z)\p_{\s}\z^r\p^{\s}\z^s\ +\phantom{^\frac{1}{1}}
{\rm fermions}\ \right)\ ,
\eeqn
where $G_{rs}$ is a hyperk\"{a}hler metric. This defines the superconformal
field theory describing the Higgs phase. From our knowledge of
the $b_{z_k}$ and $a^i_{\ha}$ field configurations corresponding to 
(\ref{2.12}) and (\ref{2.13}) one could in principle attempt to find 
the $\z^r$ corrections to the flat metric $G_{rs}=\d_{rs}$ in (\ref{action}).
 
We end this subsection by considering the dilute gas regime. Stability 
of the D-brane bound state requires that the
energy associated with the string modes should be much smaller then all
the energy scales associated with the D-brane bound state. This gives 
the condition (a similar derivation of the dilute gas regime was
given in \cite{Mald4})
\eqn
\frac{N_{L,R}}{R_1}\ll M_{1_i}\ \left(\ll M_{3_i},
\ M_{3'_i}\ll M_{5_i}^{\phantom{\frac{1}{1}}}\right)\ ,
\label{dilute}
\eeqn
where $N_{L,R}$ are the left- and right-moving momenta carried 
by the instanton strings along the $x^1$-direction (note that, e.g.
$N_R'=N_1N_2N_R$ is the level of the right-moving sector because
$L_{eff}=2\pi R_1N_1N_2$). Condition (\ref{dilute}) gives
\eqn
r_0,\ r_n\ll r_i\tan{\th_i}\ ,
\label{2.14}
\eeqn
where we define the length scales $r_n$ and $r_0$ 
according to \cite{Horo..}
\eqn
r_n^2=r_0^2\sinh^2{\b}\ ,\ \ \ \ \ \ 
N_{L,R}=\frac{R_1^2V_4}{4g^2\a'^4}r_0^2e^{\pm2\b}\ ,
\label{2.15}
\eeqn
with $V_4$ the volume of $T^4$. The condition (\ref{2.14}) defines 
the dilute gas regime derived in \cite{MaldStro}.

\subsection{Supergravity phase}

The supergravity solution associated with our D-brane bound state is a
solution of the type IIB supergravity equations of motion. The 
corresponding bosonic action is
\eqna
S_{IIB}&=&\frac{1}{2\kappa_{10}^2} \left\{\int d^{10}x \sqrt{-g}
\left [ e^{-2\phi_{10}}\left(R+4(\nabla \phi_{10})^2
-\frac{1}{2.3!} {\cal H}^2\right) -\frac{1}{2}(\partial \chi)^2 
\right.\right. \nonumber\\\label{2.16}\\
&& \ \ \ \ \left.\left. - \frac{1}{2.3!}({\cal F}_3 - \chi{\cal H})^2 
- \frac{1}{4.5!}{\cal F}_5'^2 \right ]-\frac{1}{2} \int
{\cal A}_4 \wedge {\cal H} \wedge {\cal F}_3 \right\}\ ,\nonumber
\eeqna
where $\kappa_{10}$ is the ten-dimensional gravitational coupling,
${\cal F}_5'=d{\cal A}_4+\frac{1}{2}({\cal B} \wedge {\cal F}_3 
-{\cal A}_2\wedge {\cal H})$ is a self-dual 5-form, ${\cal H}=d{\cal B}$
and ${\cal F}_3 = d{\cal A}_2$. The fields $\chi$, ${\cal A}_{2}$ and
${\cal A}_{4}$ are the 0-, 2- and 4-form 
${\rm R}\otimes {\rm R}$ potentials and the field
${\cal B}$ the 2-form ${\rm NS}\otimes{\rm NS}$
potential. $\phi_{10}$ is the dilaton
field with its zero mode subtracted. The 
${\rm NS}\otimes{\rm NS}$ background fields
describing our bound state are
\eqna
ds^{2}&=& H^{\frac{1}{2}}\left[ H^{-1}
\left( -dt^2+dx_1^2\right)
+\tilde{H}^{-1}_{\phantom{\frac{1}{1}}}\left( dx_2^2+...+dx_5^2\right)+
ds^2\left(\bE^4\right)\right]\ ,
\nonumber\\\nonumber\\
e^{2\phi}&=&H\tilde{H}^{-2}\ ,
\label{2.17}\\\nonumber\\
{\cal B}&=&-\frac{\tilde{H}^{-1}}{r^2}
\sum_ir_i^2\sin{\th_i}\cos{\th_i}
\left(dx_2^{\phantom 1}\wedge dx_3+dx_4\wedge dx_5\right)\ ,
\nonumber
\eeqna
where
\eqna
&&H=1+\frac{r_1^2+r_2^2}{r^2}+
\frac{(r_1r_2\sin{\D\th})^2}{r^4}\ ,
\nonumber\\\label{2.18}\\
&&\tilde{H}=1+\frac{r_1^2\cos^2{\th_1}+r_2^2\cos^2{\th_2}}{r^2}\ ,
\nonumber
\eeqna
with $r$ the radial coordinate on $\bE^4$.
The constants $\th_i$ and $r_i$ are defined in (\ref{2.2}) and (\ref{2.3}),
respectively. The exact form of the 
${\rm R}\otimes {\rm R}$ fields is rather complicated 
because the Chern-Simons terms for this solution do not vanish. 
We write all non-vanishing components of the 
${\rm R}\otimes {\rm R}$ fields
keeping only the corresponding leading order terms at 
infinity. The result is 
\eqna
{\cal F}_{at1}&\sim&d\left( \frac{1}{r^2}\right)_a
\sum_ir_i^2\sin^2{\th_i}+{\cal O}\left(\frac{1}{r^5}\right)\ , 
\nonumber\\\nonumber\\
{\cal F}_{abc}&\sim&-\star d\left(\frac{1}{r^2}\right)_{abc}
\sum_ir_i^2\cos^2{\th_i}+{\cal O}\left(\frac{1}{r^2}\right)\ ,
\nonumber\\\label{2.17a}\\
{\cal F}_{at123}={\cal F}_{at145}&\sim&
d\left(\frac{1}{r^2}\right)_a
\sum_ir_i^2\sin{\th_i}\cos{\th_i}
+{\cal O}\left(\frac{1}{r^5}\right)\ , 
\nonumber\\\nonumber\\
{\cal F}_{abc23}={\cal F}_{abc45}&\sim&
-\star d\left(\frac{1}{r^2}\right)_{abc}
\sum_ir_i^2\sin{\th_i}\cos{\th_i}
+{\cal O}\left(\frac{1}{r^2}\right)\ ,
\nonumber
\eeqna
where $\star$ is the dual operation with respect to the Euclidean 
metric on $\bE^4$.
This solution corresponds to the vacuum state of our D-brane bound state.

Next we obtain the D5-D1 brane solution as a special case.
All we have to do is to require that the D-3-brane charges vanish, i.e.
\eqn
r_1^2\cos{\th_1}\sin{\th_1}+r_2^2\cos{\th_2}\sin{\th_2}=0\ ,
\eeqn
and redefine the parameters $r_5$ and $r_1$ as
\eqna
&&r_5^2\equiv r_1^2\cos^2{\th_1}+r_2^2\cos^2{\th_2}\ ,
\nonumber\\\label{parameters}\\
&&r_1^2\equiv r_1^2\sin^2{\th_1}+r_2^2\sin^2{\th_2}\ .
\nonumber
\eeqna
We have then that $H=H_1H_5$, $\tilde{H}=H_5$ (note that 
$r_1r_2\sin{\D\th}\equiv r_5r_1$) and the resulting solution simplifies
dramatically (specially the ${\rm R}\otimes {\rm R}$ fields) 
to the well known D5-D1 solution.
Note that by taking $\th_1=0$ and $\th_2=\pi/2$ we also obtain the D5-D1
solution. However, our field theory description does not
hold because the gauge field diverges. In this case the correct
description is given by the D-5-brane/D-string picture \cite{CallMald}.

We may add some momentum along the string direction in (\ref{2.17}). In
the D-brane picture this corresponds to excite the left- and right-moving
sectors of the instanton strings theory. If we keep in the dilute gas
region defined in (\ref{2.14}) and further assume that 
\eqn
r_0^2,\ r_n^2\ll r_1r_2\sin{\D\th}\ ,
\label{neglmass}
\eeqn
then all the fields in (\ref{2.17}), (\ref{2.17a}) remain
unchanged but the metric which becomes
\eqna
ds^{2}&=& H^{\frac{1}{2}}\left[ H^{-1}
\left( -dt^2+dx_1^2 +\left(\frac{r_0}{r}\right)^2
\biggl(\cosh{\b}\ dt-\sinh{\b}\ dx_1\biggr)^2\right)
\right.\nonumber\\\label{2.19}\\
&&\left.\ \ \ \ \ 
+\tilde{H}^{-1}_{\phantom{\frac{1}{1}}}\left( dx_2^2+...+dx_5^2\right)
+\left(1-\left(\frac{r_0}{r}\right)^2\right)^{-1}dr^2+r^2d\O_3^2\right]\ ,
\nonumber
\eeqna
where $r_0$, $r_n$ and $\b$ are defined in (\ref{2.15}). Note that in the
case of the D5-D1 system the condition (\ref{neglmass}) follows from
(\ref{2.14}) and (\ref{2.6}). For given values of
$r_n$ and $r_0$ the total left- and right-moving momenta along the strings
are completely fixed. This means that the state of the instanton strings
is described by the microcanonical ensemble. Using the asymptotic density
of states for a conformal field theory with $4n_L\bar{n}_L$ species of
bosons and fermions we obtain the usual matching with the Bekenstein-Hawking
entropy
\eqn
S_{BH}=\frac{A_3}{4G_N^{(5)}}\left( r_nr_1r_2\sin{\D\th}\right)=
2\pi \left(\sqrt{N_L}+\sqrt{N_R}\right)\sqrt{N_1N_2n_L\bar{n}_L}\ .
\label{2.20}
\eeqn
This agreement occurs for $N_{L,R}N_1N_2\gg n_L\bar{n}_L$.
This fact may be interpreted in the following way.
We may approximate the microcanonical ensemble by the canonical ensemble
with the left- and right-moving temperatures \cite{MaldStro}
\eqn
T_{L,R}=\frac{1}{\pi R_1}\sqrt{\frac{N_{L,R}}{N_1N_2n_L\bar{n}_L}}=
\frac{r_0 e^{\pm\b}}{2\pi r_1r_2\sin{\D\th}}\ .
\label{2.21}
\eeqn
The occupation number for a given mode is then easily calculated in
the canonical ensemble. This approximation is valid for $T_{L,R}\gg E_g$,
where $E_g\sim (R_1N_1N_2)^{-1}$ is the energy gap on the field theory
side. Physically this means that in the thermodynamical description the
energy spectrum may be regarded as continuous. Replacing for the values 
of $T_{L,R}$ we obtain precisely the condition 
$N_{L,R}N_1N_2\gg n_L\bar{n}_L$. Thus, the non-extreme case (\ref{2.19}) 
is associated with a thermal state of the instanton strings.

Now we comment on the region of validity of the supergravity approximation.
We keep $g\ll 1$ in order to suppress closed string loop effects. We are
also making an $\a'$ expansion. Thus, the length scales in our solution
have to be much larger then one in string units, i.e.
\eqn
r_0,\ r_n,\ r_i\sin{\th_i},\ \sqrt{r_1r_2\sin{\D\th}}\gg 1\ .
\label{2.22}
\eeqn
The supergravity approximation is valid for processes involving energy 
scales such that $\o l_{max}\ll 1$ where $l_{max}$ is the maximal
length scale \cite{Mald3}.
We conclude that the D-brane and supergravity
phases are mutually exclusive. Considering the last condition in 
(\ref{2.22}) for the region of validity of the supergravity phase
we have in terms of the D-brane system 
$g\sqrt{N_1N_2n_L\bar{n}_L}\gg 1$. We shall see below that
consistency between the supergravity and D-brane phases requires 
$n_L\bar{n}_L$ not to be very large. Hence we have (for $N_1\sim N_2$)
\eqn
gN_i\gg 1\ .
\label{largeN}
\eeqn
Since $g$ is small we conclude that $N_i\gg 1$. Thus, the supergravity
phase is associated with a large $N_i$ D-brane system.

Next we show that to compare with the supergravity phase it is perfectly 
consistent to neglect the massive string states on the field theory
side. This corresponds to the $\a'\rightarrow 0$ decoupling limit where
these states become infinitely massive. The condition to neglect 
such modes is
\eqn
T_{L,R}\ll\frac{1}{\sqrt{\a'}}\ \ \Leftrightarrow\ \ 
r_n^2,\ r_0^2\ll\left(r_1r_2\sin{\D\th}\right)^2\frac{1}{\a'}\ .
\label{2.23}
\eeqn
Using the conditions (\ref{neglmass}) and (\ref{2.22}) it is seen that
(\ref{2.23}) holds. Thus, on the supergravity side we do not expect to 
find effects caused by these fields and it is consistent to drop them in
the field theory approach (note that in this case we are not protected 
by supersymmetry as it was the case in \cite{CostaPerry}).

Another check of consistency between both descriptions is concerned with
the mass gap. In the field theory description this equals $(N_1N_2R_1)^{-1}$, 
while on the supergravity side it is given by the inverse of the 
temperature such that the specific heat is of order unit
\cite{MaldSuss,Pres..,HolzWilc,KrausWilc}. This condition
gives
\eqn
\d M\sim \frac{G_N^{(5)}}{(r_1r_2\sin{\D\th})^2}
\sim (N_1N_2n_L\bar{n}_LR_1)^{-1}\ .
\label{massgap}
\eeqn
Thus, we can not have $n_L\bar{n}_L$ very big. For the D5-D1 brane system  
this fact brings us to the case where the DBI corrections are important
that we have discussed in subsection 2.1.3. 
In the more general case we may have $n_L\bar{n}_L\sim 1$ while
keeping $p_i\gg |q_i|$ and $\bar{p}_i\gg |\bar{q}_i|$ (for example this
happens for $q_i=\bar{q}_i=1$ and $p_1=p_2-1$, $\bar{p}_1=\bar{p}_2-1$
while keeping $p_i\gg 1$ and $\bar{p}_i\gg 1$).

\section{Minimally coupled scalar}
\news

In this section we shall find a minimally coupled scalar in the
supergravity backgrounds of section 2.2. We shall follow the same strategy 
of \cite{Call..} by reducing the type IIB action to five-dimensions. Then
we linearise the DBI action and generalise the result to the non-abelian
case in order to determine the coupling of the minimally coupled scalar
to the instanton strings.

\subsection{Reduction to five dimensions}

To find a minimally coupled scalar in our black hole backgrounds we 
reduce the action (\ref{2.16}) with the following metric ansatz \cite{Call..}
\eqn
ds^2=e^{\frac{4}{3}\phi_5}g_{ab}dx^adx^b+
e^{2\n_5}\left( dx^1+{\cal A}_a^{(K)}dx^a\right)^2+
e^{2\n}\d_{\ha\hb}dx^{\ha}dx^{\hb}\ ,
\label{3.1}
\eeqn
where $g_{ab}$ is the five-dimensional Einstein metric\footnote{In this 
subsection the indices $a,b,...$ run over 0,6,...9, otherwise they are
ten-dimensional spacetime indices.}. Truncating the
action (\ref{2.16}) such that the only non-vanishing form fields are those
appearing in the solution (\ref{2.17}), (\ref{2.17a}) and assuming as it
is the case that ${\cal A}_{a 1}$, ${\cal A}_{a 1\ha\hb}$ and 
${\cal A}_a^{(K)}$ are electric we obtain the following 
five-dimensional action ($S_1$ and $S_2$ were vanishing in the case 
considered in \cite{Call..})
\eqna
S&=&\frac{1}{2\kappa_5^2}\int d^5x \sqrt{-g}\left[
R-(\p\Phi)^2-\frac{4}{3}(\p\la)^2-4(\p\n)^2
-\frac{1}{4}e^{\frac{8}{3}\la}\left({\cal F}_{ab}^{(K)}\right)^2
\right.\nonumber\\\nonumber\\
&&\ \ \ \ \ \ \left.
-\frac{1}{2.2!}e^{-\frac{4}{3}\la+4\n}\left({\cal F}_{ab 1}\right)^2
-\frac{1}{2.3!}e^{\frac{4}{3}\la+4\n}\left({\cal F}_{abc}\right)^2
\right]+S_1+S_2\ ,
\nonumber\\\nonumber\\
S_1&=&\frac{1}{2\kappa_5^2}\int d^5x \sqrt{-g}\left[
-\frac{1}{2.2!}e^{-4\nu}\left(\p_{a}{\cal B}_{\ha\hb}\right)^2\right.
\label{3.2}\\\nonumber\\
&&\ \ \ \ \ \ \left.
-\frac{1}{4.2!2!}e^{-\frac{4}{3}\la}\left({\cal F'}_{ab 1\ha\hb}\right)^2
-\frac{1}{4.2!3!}e^{\frac{4}{3}\la}\left({\cal F'}_{abc\ha\hb}\right)^2
\right]\ ,
\nonumber\\\nonumber\\
S_2&=&\frac{1}{4\kappa_5^2}\int d^5x\frac{1}{2!2!}
\e^{\ha\hb\hth\hg}\e^{abcde}\left(
\frac{1}{3!}{\cal A}_{a1\ha\hb}{\cal F}_{bcd}{\cal H}_{e\hth\hg}
-\frac{1}{2!2!}{\cal A}_{ab\ha\hb}{\cal F}_{cd1}{\cal H}_{e\hth\hg}
\right)\ ,
\nonumber
\eeqna
where $\kappa_5$ is the five-dimensional gravitational coupling and 
\eqn
\Phi=\phi_{10}-2\n=\phi_5+\frac{\n_5}{2}\ ,\ \ \ \ 
\la=\n_5-\frac{\Phi}{2}=\frac{3}{4}\n_5-\frac{\phi_5}{2}\ .
\label{3.3}
\eeqn
The 5-form ${\cal F}'_5$ reduces to
\eqna
&&{\cal F}'_{ab1\ha\hb}={\cal F}_{ab1\ha\hb}+\frac{1}{2}
\left({\cal B}_{\ha\hb}{\cal F}_{ab1}
-2{\cal A}_{1[a}{\cal H}_{b]\ha\hb}\right)\ ,
\nonumber\\\label{3.4}\\
&&{\cal F}'_{abc\ha\hb}={\cal F}_{abc\ha\hb}+\frac{1}{2}
\left({\cal B}_{\ha\hb}{\cal F}_{abc}
-3{\cal A}_{[ab}{\cal H}_{c]\ha\hb}\right)\ ,
\nonumber
\eeqna
and the ten-dimensional self-duality condition $\star{\cal F'}={\cal F'}$
becomes
\eqn
{\cal F}'_{abc\ha\hb}=\frac{1}{2!2!}\sqrt{-g}e^{-\frac{4}{3}\la}
\e_{abcde}\e_{\ha\hb\hth\hg}{\cal F'}^{de}_{\ \ 1\hth\hg}\ .
\label{3.5}
\eeqn
We conclude that the field $\Phi$ (dilaton field in the six-dimensional 
theory) is minimally coupled.

\subsection{Coupling to the instanton strings}

The coupling of the scalar field $\Phi$ to the instanton strings may be 
found following a similar approach to the D-3-brane case \cite{Kleb,Kleb..}.
Start with the DBI action for the D-5-brane written in the static gauge
\eqna
&&S_{DBI}=-T_5\int d^6x\ e^{-\phi_{10}}
\sqrt{-\det{\left(\hat{g}+{\cal G}\right)}}\ 
+{\rm \ RR\ couplings}\ ,
\nonumber\\\nonumber\\
&&{\cal G}_{\a\b}=2\pi\a'G_{\a\b}-\hat{{\cal B}}_{\a\b}\ ,
\label{3.6}\\\nonumber\\
&&\hat{g}_{\a\b}=
g_{\a\b}+2g_{m(\a}\p_{\b)}X^m+g_{mn}\p_{\a}X^m\p_{\b}X^n\ .
\nonumber
\eeqna
As for $\hat{g}_{\a\b}$ the field $\hat{{\cal B}}_{\a\b}$ is the 
pull-back to the D-5-brane worldvolume of the 
${\rm NS}\otimes{\rm NS}$ 2-form potential.  
We set ${\cal B}$ to zero and expand the
metric around flat space: $g_{ab}=\eta_{ab}+h_{ab}$. Then we expand the
action (\ref{3.6}) keeping the quadratic terms in the worldvolume fields
and the linear terms in the bulk fields. Defining the scalar fields
$\phi^m=X^m/(2\pi\a')$ the result is 
\eqna
S_{DBI}&\sim&-(2\pi\a')^2T_5\int d^6x\left[
\left(1-\phi_{10}\right)
\left(\frac{1}{4}\left(G_{\a\b}\right)^2
+\frac{1}{2}\p_{\a}\phi^m\p^{\a}\phi_m\right)
\right.
\nonumber\\\nonumber\\
&&\ \ \ \ \ \ \ \ \ \ \ \ \ \ \ \ \ \ \ \ \ \ \ \ \left.
-\frac{1}{2}h^{\a\b}T_{\a\b}
+\frac{1}{2}h_{mn}\p_{\a}\phi^m\p^{\a}\phi^n\right]\ ,
\label{3.7}\\\nonumber\\
T_{\a\b}&=&G_{\a}^{\ \th}G_{\b\th}
-\frac{1}{4}\eta_{\a\b}\left(G_{\th\gamma}\right)^2
+\p_{\a}\phi^m\p_{\b}\phi_{m}
-\frac{1}{2}\eta_{\a\b}\p_{\th}\phi^m\p^{\th}\phi_m\ ,
\nonumber
\eeqna
where the indices are raised and lowered with respect to the Minkowski
metric and $T_{\a\b}$ is the energy-momentum tensor of the abelian
YM action (free terms in (\ref{3.7})). The coupling 
between the fields $\Phi$ and $B_{\a}$ is determined by the coupling
of $\phi_{10}$ and $h^{\a\b}$ to $B_{\a}$.
Therefore we drop the last term in the action (\ref{3.7}). The obvious
generalisation of the interacting action to the $U(N)$ case is
\eqna
S_{int}&=&\frac{1}{g^2_{YM}}\int d^6x\left[
\phi_{10}{\rm tr}\left(\frac{1}{4}\left(G_{\a\b}\right)^2+
\frac{1}{2}\left(\p_{\a}\phi_m+i[B_{\a},\phi_m]\right)^2
-\frac{1}{4}[\phi_m,\phi_n]^2\right)+\frac{1}{2}h^{\a\b}T_{\a\b}\right],
\nonumber\\\nonumber\\
T_{\a\b}&=&{\rm tr}\left( G_{\a}^{\ \th}G_{\b\th}
-\frac{1}{4}\eta_{\a\b}\left(G_{\th\gamma}\right)^2
+(\p_{\a}\phi^m+i[B_{\a},\phi^m])(\p_{\b}\phi_m+i[B_{\b},\phi_m])\right.
\label{3.8}\\\nonumber\\
&&\left.\ \ \ \ 
-\frac{1}{2}\eta_{\a\b}(\p_{\th}\phi_m+i[B_{\th},\phi_m])^2
+\frac{1}{4}\eta_{\a\b}[\phi_m,\phi_n]^2\right)\ .
\nonumber
\eeqna
The situation is analogous to the calculation involving the D-3-brane
\cite{Kleb,Kleb..}. Note that it is
straightforward to write the supersymmetric completion of (\ref{3.8})
because $\phi_{10}$ couples to the SYM action and $h^{\a\b}$ to the
corresponding energy-momentum tensor. We are just writing the interacting
terms that follow from the SYM action but there will be DBI type
corrections as well as there may be modifications to the energy-momentum
tensor imposed by conformal invariance \cite{KlebGubs}.

In the linear approximation the scalar fields in the ansatz (\ref{3.1})
are identified with the tensor $h_{\a\b}$ according to
\eqn
h=h^{\ha}_{\ \ha}=8\n\ ,\ \ \ h_{00}=-\frac{4}{3}\phi_{5}\ ,\ \ \ 
h_{11}=2\n_5\ .
\label{3.9}
\eeqn
Keeping only the interacting terms with the field $\Phi$ that are 
quadratic in the worldvolume fluctuating field $A_{\a}$ we have 
(note that $G_{\a\b}=G^0_{\a\b}+F_{\a\b}$)
\eqn
S_{int}=\frac{1}{g^2_{YM}}\int d^6x\ \Phi\ {\rm tr}\left(
\frac{1}{2}F_{\s\ha}F^{\s\ha}\right)\ .
\label{3.10}
\eeqn
As explained before we are just considering processes involving the
massless excitations on the brane and keeping only the fields associated
with the instanton strings. A similar calculation to the one
in subsection 2.1.3 gives the following interacting term between $\Phi$
and the instanton strings
\eqn
S_{int}=-\frac{T_{ins}}{2}\int dt\int_0^{L_{eff}}dx^1\ \Phi\ 
\sum_{r=1}^{4n_L\bar{n}_L}\p_{\s}\z^r\p^{\s}\z^r\ ,
\label{3.11}
\eeqn
where $T_{ins}$ and $L_{eff}$ are given in (\ref{tension}). The
factor multiplying the integral is important.

\section{Cross section}
\news

In this section we calculate the cross section for the scattering of the
D-brane bound state by the scalar particle $\Phi$ both in the supergravity
and D-brane picture. We shall consider the supergravity solutions
corresponding to the D-brane system vacuum and thermal states.

\subsection{Classical calculation}

The equation of motion satisfied by the scalar field in the 
background (\ref{2.17}) is for the s-wave mode
\eqn
\left[r^{-3}\frac{d}{dr}r^3\frac{d}{dr}
+\o^2\left( 1+\frac{r_1^2+r_2^2}{r^2}
+\frac{(r_1r_2\sin{\D\th})^2}{r^4}\right)\right]\Phi(r)=0\ .
\label{4.1}
\eeqn
Writing $\Phi=\rho^{-3/2}\Psi(\rho)$ where $\rho=\o r$ we have
\eqn
\left[\frac{d^2}{d\rho^2}+1
-\frac{3/4-\o^2\left( r_1^2+r_2^2\right)}{\rho^2}
+\frac{\o^4(r_1r_2\sin{\D\th})^2}{\rho^4}\right]\Psi(\rho)=0\ .
\label{4.2}
\eeqn
For $\rho\gg\o r_i\sin{\D\th}$ (i.e. $r\gg r_i\sin{\D\th}$) we may 
neglect the ${\cal O}(1/\rho^4)$ term in comparison with the 
${\cal O}(1/\rho^2)$ terms.
In the low energy limit $\o r_i\ll 1$ we are considering, the differential
equation satisfied by $\Psi(\rho)$ becomes
\eqn
\left[\frac{d^2}{d\rho^2}+1
-\frac{3}{4\rho^2}\right]\Psi(\rho)=0\ ,
\label{4.3}
\eeqn
which is solved in terms of Bessel functions of degree 
one.\footnote{An alternative resolution is to keep the $\o r_i$ terms, 
solve the differential equation in terms of Bessel functions of degree 
$\pm\sqrt{1-\o^2\left( r_1^2+r_2^2\right)}$ and take the limit 
$\o r_i\ll 1$ at the end. Within our approximation the final result 
is the same.}

If we perform instead the coordinate transformation
$y=\o r_1r_2\sin{\D\th}/(2r^2)$ the differential equation (\ref{4.1}) 
becomes
\eqn
\left[\frac{d^2}{dy^2}+\o\frac{r_1r_2\sin{\D\th}}{2y}
+\o^2\frac{r_1^2+r_2^2}{4y^2}
+\o^3\frac{r_1r_2\sin{\D\th}}{8y^3}\right]\Phi(y)=0\ .
\label{4.4}
\eeqn
The last term may be neglected for $y\gg\o \sin{\D\th}$ (i.e. $r\ll r_i$). 
In the coordinate $z=\sqrt{2y\o r_1r_2\sin{\D\th}}$ and 
with $\Phi=z\Upsilon(z)$ we have in the limit $\o r_i\ll 1$
\eqn
\left[ z^2\frac{d^2}{dz^2}+z\frac{d}{dz}+(z^2-1)\right]\Upsilon(z)=0\ ,
\label{4.5}
\eeqn
which is again solved in terms of Bessel functions of degree one. 
Since $\D\th\ll 1$ we conclude that both the equations (\ref{4.3}) and
(\ref{4.5}) have a large overlapping domain and the corresponding
solutions may be patched together.\footnote{If $\Delta\th\sim1$ which 
happens for the D5-D1 brane system (with $f=16$) the near and far regions 
do not overlap. In this case there are $\omega r_i$ corrections which 
are suppressed within our approximation \cite{MaldStro}.}

The cross section may be calculated by using the flux method 
\cite{MaldStro}. In the near zone we require a purely infalling 
solution at $r=0$ and match it to the solution in the far zone.
The result is
\eqna
{\rm Near\ region:} & \Phi=z\biggl( J_1(z)+N_1(z)\biggr)\ ,
\nonumber\\\label{4.6}\\
{\rm Far\ region:} & \Phi=-{\displaystyle \frac{4}{\pi\rho}}J_1(\rho)\ ,
\nonumber
\eeqna
where $J_1$ and $N_1$ are Bessel and Neumann functions, respectively. 
The cross section is obtained from the ratio between the
flux at the horizon and the incoming flux at infinity
\eqn
\sigma_{abs}=\frac{4\pi}{\o^3}\frac{{\cal F}_h}{{\cal F}^{inc}_{\infty}}=
\pi^3\o \left( r_1r_2\sin{\D\th}\right)^2\ .
\label{4.7}
\eeqn

The calculation of the cross section for the non-extreme case is
similar to the calculation presented in \cite{MaldStro}. In the far
region the solution is the same as in the previous case and in the near
region $\Phi$ is expressed in terms of hypergeometric functions. The
result is
\eqn
\sigma=A_h\frac{\pi\o}{2}
\frac{e^{\frac{\o}{T_H}}-1}
{\left(e^{\frac{\o}{2T_L}}-1\right)\left(e^{\frac{\o}{2T_R}}-1\right)}\ ,
\label{4.8}
\eeqn
where $A_h$ is the horizon area, the left- and right-moving
temperatures were defined in (\ref{2.21}) and
\eqn
\frac{1}{T_H}=\frac{1}{2}\left(\frac{1}{T_L}+\frac{1}{T_R}\right)\ ,
\label{4.9}
\eeqn
is the inverse of Hawking temperature. 

\subsection{D-brane calculation}

Now we calculate the absorption probability for incoming scalar
particles when the D-brane system is in the vacuum state. The canonically
normalised fields are $\tilde{\Phi}=\Phi/\kappa_6$ and 
$\tilde{\z}^r=\sqrt{T_{ins}}\z^r$. These fields have the following mode
expansion
\eqna
\tilde{\z}^r&=&\sum_q\frac{1}{\sqrt{2L_{eff}q_0}}
\left(\z^r_qe^{iq_{\s}x^{\s}}+
\z_q^{r\dagger}e^{-iq_{\s}x^{\s}}\right)\ ,
\nonumber\\\label{4.10}\\
\tilde{\Phi}&=&\sum_{k_1,\vec{k}}
\frac{1}{\sqrt{2L_1{\cal V}_4k_0}}
\left(\Phi_ke^{ik\cdot x}+\Phi_k^{\dagger}e^{-ik\cdot x}\right)\ ,
\nonumber
\eeqna
where $q$ and $k_1$ are the corresponding momenta along the string 
direction and $\vec{k}$ the momentum in the transverse 
space with volume ${\cal V}_4$. Note that we are considering a 
six-dimensional free action for the field $\Phi$ that arises from 
compactification of the IIB theory on $T^4$ and not the 
five-dimensional action (\ref{3.2}). The dependence on the
string direction will in fact be irrelevant because we shall consider modes
of the field satisfying $k_1=0$, i.e. we do not consider charged particles
\cite{MaldStro,KlebGubs1}.
However, it is important to realize that the scalar particle is defined
on a length $L_1=2\pi R_1$ while the instanton strings on a length
$L_{eff}=L_1N_1N_2$. We have normalised the states such that 
$|q\rangle=\z_q^{r\dagger}|0\rangle$ represents a single particle with
momentum $q$ in the length $L_{eff}$ and 
$|k\rangle=\Phi_k^{\dagger}|0\rangle$ a single particle with momentum
$k$ in the volume $L_1{\cal V}_4$. Thus, from the spacetime perspective (i.e.
integrating over the string direction) a state $|k\rangle$ carries a flux
$1/{\cal V}_4$.

In terms of the canonically normalised fields the interacting vertex 
(\ref{3.11}) becomes
\eqn
S_{int}=\frac{\kappa_6}{2}\int dt\int_0^{L_{eff}}dx^1\tilde{\Phi}
\sum_{r=1}^{4n_L\bar{n}_L}\p_{\s}\tilde{\z}^r\p^{\s}\tilde{\z}^r\ .
\label{4.11}
\eeqn
The initial and final states for the process considered are
\eqna
|i\rangle=\Phi_k^{\dagger}|0\rangle\ , & k=(k_0,0,\vec{k})\ ,
\nonumber\\\label{4.12}\\
|f\rangle=\z_q^{r\dagger}\z_p^{r\dagger}|0\rangle\ , & 
q=(q_0,q_1)\ ,\ \ p=(p_0,p_1)\ .
\nonumber
\eeqna
The amplitude for this process is then
\eqn
T_{fi}=-\frac{\kappa_6}{2}
\frac{p\cdot q}
{\sqrt{2L_1{\cal V}_4k_0}\sqrt{2L_{eff}q_0}\sqrt{2L_{eff}p_0}}2\ .
\label{4.13}
\eeqn
The reason the supersymmetric completion of (\ref{4.11}) was not 
considered is that on-shell fermions give a vanishing 
contribution to this amplitude.
The final factor of two is because either of the $\z$'s in (\ref{4.11})
may annihilate either of the final particle states in (\ref{4.12})
\cite{DasMath}. The probability per unit of time for this transition 
to occur is then
\eqn
\Gamma_{fi}=L_{eff}(2\pi)^2\d(k_0-p_0-q_0)\d(p_1+q_1)|T_{fi}|^2\ .
\label{4.14}
\eeqn
To obtain the total probability rate we have to sum over the 
$4n_L\bar{n}_L$ species of particles and integrate over the final
momenta dividing by two due to particle identity. The result is
\eqn
\Gamma_{abs}=4n_L\bar{n}_L\frac{1}{2}\sum_{p,q}\Gamma_{fi}=
\frac{1}{4{\cal V}_4}L_{eff}n_L\bar{n}_L\kappa_5^2\o\ .
\label{4.15}
\eeqn
Since the state $|i\rangle=|k\rangle$ carries the flux $1/{\cal V}_4$ we have
that $\s_{abs}={\cal V}_4\Gamma_{abs}$ which agrees exactly with (\ref{4.7}).
Besides the rather successfully D-3-brane case \cite{Kleb,Kleb..,KlebGubs} 
this is the first example
where this calculation has been done by deducing from first principles
the coupling between the bulk and the worldvolume fields.

The calculation when the D-brane system is described by a thermal state
of left- and right-movers is done in the following way. We consider a 
unit normalised state $|n_R,n_L\rangle$ of the instanton strings 
with $n(p_R)$ and $n(p_L)$ right- and left-mover occupation numbers. 
Now the initial and final states for the process are
\eqna
|i\rangle=\Phi_k^{\dagger}|n_R,n_L\rangle\ , & k=(k_0,0,\vec{k})\ ,
\nonumber\\\label{4.16}\\
|f\rangle=\z_q^{r\dagger}\z_p^{r\dagger}|n_R,n_L\rangle\ , & 
q=(q_0,q_1)\ ,\ \ p=(p_0,p_1)\ .
\nonumber
\eeqna
The amplitude for this process is
\eqn
T_{fi}=-\frac{\kappa_6}{2}
\frac{p\cdot q}{\sqrt{2L_1{\cal V}_4k_0}\sqrt{2L_{eff}q_0}\sqrt{2L_{eff}p_0}}2
(n(p)+1)(n(q)+1)\ .
\label{4.17}
\eeqn
The total probability rate is obtained by summing over all final states
and averaging over all initial states in the thermal ensemble 
\cite{Dhar..}. This
gives the desirable Bose-Einstein thermal factors. Agreement with 
(\ref{4.8}) is found by using $\s_{abs}=V_4(\Gamma_{abs}-\Gamma_{emis})$,
where $\Gamma_{emis}$ is the probability rate for the time reversed process
\cite{Call..}.

\section{CFT/AdS duality}
\news

In this section we start by analysing the region of validity of the
previous cross section calculations. We shall define a double scaling
limit \cite{Kleb} where the supergravity cross section calculation 
and our gauge theory calculation of the D-brane absorption probability
should in fact agree. 
The last subsections are concerned with the near horizon 
geometry and Maldacena's duality proposal \cite{Mald2}.

\subsection{Double scaling limit}

Consider the ground state of the D-brane system. We argued in subsection
2.2 that the supergravity approximation holds if the length scales in the
solution are big in string units. In particular we have
\eqn
r_1r_2\sin{\D\th}\gg 1\ \ 
\Rightarrow\ \ g_{eff}\sin{\D\th}\gg 1\ ,
\label{5.1}
\eeqn
where $g_{eff}$ is the D-brane effective string coupling defined 
in (\ref{2.3}). In this limit string
corrections to the metric are suppressed (the string loop corrections
are also suppressed because we are considering $g\ll 1$). The curvature
of this background is bounded by its value at $r=0$ where
\eqn
{\rm curv}\sim 
\frac{1}{r_1r_2\sin{\D\th}}\sim \frac{1}{g_{eff}\a'\sin{\D\th}}\ .
\label{5.2}
\eeqn
The classical cross section is naturally expanded in powers
of $\o^4{\rm curv}^{-2}$. Thus, for energies such that
\eqn
\o^4(g_{eff}\a'\sin{\D\th})^2\ll 1\ ,
\label{5.3}
\eeqn
we expect the classical approximation to the scattering process to be good.
Both conditions (\ref{5.1}) and (\ref{5.3}) are satisfied in the double
scaling limit \cite{Kleb}
\eqn
g_{eff}\sin{\D\th}\rightarrow\infty\ ,\ \ \ \ 
\o^4\a'^2\rightarrow  0,
\label{5.4}
\eeqn
such that
\eqn
\left(g_{eff}\sin{\D\th}\right)^2
\left(\o^4\a'^2\right)\ ,
\label{5.5}
\eeqn
is held fixed and small, i.e. (\ref{5.3}) holds.
The second condition in (\ref{5.4}) implies that the massive
excitations on the D-5-brane may be neglected when comparing with the
supergravity cross section calculation, i.e. it corresponds to the
decoupling limit of the brane theory. These states have a mass
that scales as $1/\sqrt{\a'}\gg \o$ (note that some of the massive
states have a mass proportional to $\sqrt{\D\th}$ which is held fixed 
in the limit (\ref{5.4})).
This is the reason why they were dropped in the field theory
description. They may be neglected in the double scaling limit.

Now we show that in the limit (\ref{5.4}) the D-brane calculation may in
fact be trusted (even if $g_{eff}\rightarrow\infty$). 
The only scale in the
scattering calculation is given by the gravitational coupling $\kappa_6$ as
may be seen in the interacting Lagrangian when written in terms
of the canonically normalised fields. The cross section is then
an expansion in powers of
\eqn
\o^4\kappa_6^2n_L\bar{n}_L\frac{L_{eff}}{L_1}\ .
\label{5.6}
\eeqn
The $n_L\bar{n}_L$ factor is because we sum over all different species
in the final state and the $L_{eff}/L_1=N_1N_2$ factor because the scalar
particles leave in a length $L_1$ while the instanton strings in a length
$L_{eff}$ (the state $|k\rangle=\Phi_k^{\dagger}|0\rangle$ corresponds
to a single particle in the volume $L_1{\cal V}_4$ or $L_{eff}/L_1$
particles in the volume $L_{eff}{\cal V}_4$). The $\o^4$ factor 
follows from dimensional analysis. Thus, the perturbative string 
calculation is valid for
\eqn
\o^4N_1N_2n_L\bar{n}_L\kappa_6^2\ll 1\ \ \Rightarrow\ \ 
\o^4\left(\a'g_{eff}\sin{\D\th}\right)^2\ll 1\ ,
\label{5.7}
\eeqn
which holds in the limit (\ref{5.4}).

We conclude that both the classical and string cross section calculations
have an overlapping domain of validity and it is therefore not 
surprising that agreement is found.

\subsection{Near horizon geometry}

As it is the case for the D5-D1 brane configuration \cite{Hyun,Boon..,SfetSken}
the near horizon geometry associated with our D-brane bound state is 
$AdS_3\times S^3\times M$, where $M$ is a compact manifold ($T^4$ in
our case). Taking the limit $r\rightarrow 0$ we obtained the
following fields describing the near horizon geometry
\eqna
ds_{10}^2&\sim& ds_3^2+
\frac{R^2}{r_1^2\cos^2{\th_1}+r_2^2\cos^2{\th_2}}ds^2(T^4)+R^2d\O_3^2\ ,
\nonumber\\\nonumber\\
ds_3^2&=&
-\frac{\rho^2}{R^2}d\tau^2+\frac{R^2}{\rho^2}d\rho^2+\rho^2d\varphi^2\ ,
\label{5.8}\\\nonumber\\
e^{2\phi}&\sim&\frac{R^4}{(r_1^2\cos^2{\th_1}+r_2^2\cos^2{\th_2})^2}\ ,
\nonumber\\\nonumber\\
{\cal F}_3&\sim& 2(r_1^2\cos^2{\th_1}+r_2^2\cos^2{\th_2})\e_{3}\ ,
\nonumber
\eeqna
where $R^2=r_1r_2\sin{\D\th}$, $\tau=\frac{R}{R_1}t$, 
$\varphi=\frac{x_1}{R_1}$, $\rho=\frac{R_1}{R}r$ and $\e_{3}$ is the 
unit 3-sphere volume form. To obtain the horizon value for ${\cal F}_3$
we used the behaviour at the horizon of the Chern-Simons terms in the
solution. Note that the electric terms in ${\cal F}_3$ as well as 
${\cal H}$ and ${\cal F}_5$ vanish at the horizon. $R$ is the 3-sphere 
radius and the $AdS_3$ cosmological constant is given by $\Lambda=-R^{-2}$.
This geometry is interpreted as the ${\rm R}\otimes {\rm R}$ ground state 
of string theory on the $AdS_3\times S^3\times T^4$ background 
\cite{CousHenn,Stro}.

For later convenience we express the parameters in the solution 
(\ref{5.8}) in terms of the field theory quantities
\eqna
R^2&=&g\a'\sqrt{N_1N_2n_L\bar{n}_L}\sqrt{\frac{\a'^2}{R_2...R_5}}
\sim g\a'\sqrt{N_1N_2n_L\bar{n}_L}\equiv g\a'\sqrt{Q_1Q_5}\ ,
\nonumber\\\nonumber\\
{\cal F}_3&\sim&2g\a'(N_1p_1\bar{p}_1+N_2p_2\bar{p}_2)\e_{3}
\equiv 2g\a'Q_5\e_{3}\ ,
\label{5.9}\\\nonumber\\
v_f&\equiv&\frac{V_f(T^4)}{(2\pi)^4\a'^2}=
\frac{N_1N_2n_L\bar{n}_L}{(N_1p_1\bar{p}_1+N_2p_2\bar{p}_2)^2}
\equiv \frac{Q_1}{Q_5}\ .
\nonumber
\eeqna
The last identifications in these equations correspond to the
D5-D1 brane case. The $T^4$ volume has its fixed value
at the horizon while the six-dimensional string coupling $g_6$ has
the same value as in the original solution where it was constant \cite{Mald2}.

For the non-extreme solution the three-dimensional geometry in (\ref{5.8})
valid in the near horizon region is replaced by the BTZ black hole
\cite{Bana..}. In the previous $(\tau,\varphi)$ coordinates and 
defining a new radial coordinate $\rho^2=R_1^2(r^2+r_0^2\sinh^2{\b})/R^2$ 
\cite{BalaLars} the resulting metric reads
\eqna
&&ds_3^2=-N^2d\tau^2+N^{-2}d\rho^2+
\rho^2\left(d\varphi-N_{\varphi}d\tau\right)^2\ ,
\nonumber\\\label{5.10}\\
&&N^2=\frac{\rho^2}{R^2}-\frac{R_1^2r_0^2\cosh{2\b}}{R^4}
+\frac{R_1^4r_0^4\sinh{2\b}}{4R^6\rho^2}\ ,\ \ \ 
N_{\varphi}=\frac{R_1^2r_0^2\sinh{2\b}}{2R^3\rho^2}\ .
\nonumber
\eeqna
The derivation of this metric assumes that we are in the dilute gas 
regime and that the condition to neglect the massive string states
is satisfied, i.e.
\eqn
r^2,\ r_n^2,\ r_0^2\ll r_i^2\tan^2{\th_i},\ r_1r_2\sin{\D\th}\ ,
\label{5.11}
\eeqn
where we are assuming that we are close enough to the horizon such that
$r$ satisfies these conditions.
This geometry corresponds to an excited (thermal state) of string theory
on the $AdS_3\times S^3\times T^4$ background \cite{Mald2}. Using the 
formulation of quantum gravity in $2+1$ dimensions as a topological 
Chern-Simons theory \cite{AchuTown,Witt2}, Carlip found that the 
degrees of freedom at the horizon are described by a $1+1$-dimensional 
conformal field theory reproducing the entropy formula for the BTZ 
black hole \cite{Carl}. A different approach originally due to 
Strominger \cite{Stro,BalaLars} is based on the fact 
that any quantum theory of gravity on $AdS_3$ has an asymptotic 
algebra of diffeomorphisms given by the Virasoro algebra 
\cite{CousHenn}.\footnote{See refs. [64-80] for recent work on the subject.}
Physical states will form representations of such algebra
and the correct entropy formula follows (for the correct central charge).
Of course all these results are valid in our model because the
near horizon geometry is similar to the D5-D1 brane case. The only
difference is the way we parametrise the solutions.

\subsection{CFT/AdS correspondence}

The motivation for the conjectured duality between the decoupled theory 
on the brane and supergravity on the $AdS_3\times S^3\times T^4$ background
\cite{Mald2} relies in part on the agreement between the entropy 
calculations and the scattering calculations in the double scaling limit 
(\ref{5.4}). The region of validity of the supergravity description of
the near horizon geometry is given in eqn. (\ref{5.1}) which reads
\eqn
\frac{R^2}{\a'}\sim g\sqrt{N_1N_2n_L\bar{n}_L}\gg 1\ .
\eeqn
This may be accomplished by taking the 
$\a'\rightarrow 0$ limit
\eqn
g_{YM}^2=(2\pi)^3g\a'\rightarrow 0\ ,\ \ \ 
N_1\sim N_2\rightarrow \infty\ ,
\label{5.12}
\eeqn
such that
\eqn
R^2\sim g^2_{YM}\sqrt{N_1N_2n_L\bar{n}_L}\ ,
\label{5.13}
\eeqn
is held fixed. Note that in this limit all the fields in (\ref{5.8}) are
held fixed as may be seen from (\ref{5.9}). This limit is equivalent to
the double scaling limit (\ref{5.4}), the difference is that the energies
are held fixed and $\a'\rightarrow 0$.

Now on the D-brane side the limit (\ref{5.12}) is just the 't Hooft large 
$N$ limit. The advantage of formulating Maldacena's duality conjecture 
using this model is that we know, at least in principle, 
the action for the D-5-brane and its coupling to the bulk fields. 
As in the analysis given by Alwis for the D-3-brane \cite{Alwis} 
we consider the 't Hooft scaling for the
D-5-brane action and see what conclusions we may draw. Schematically 
't Hooft scaling may be analysed by writing (the factor $\sqrt{N_1N_2}$
replaces the usual factor of $N$ because we are considering the Higgs branch 
of the theory and the fields $b_{\a}$ are in the fundamental representation
of $U(N_1p_1\bar{p}_1)\otimes U(\overline{N_2p_2\bar{p}_2})$)
\eqn
S\sim -\frac{\sqrt{N_1N_2}}{\a'^2}
\frac{\a'^2}{\sqrt{N_1N_2}g^2_{YM}}
\int d^6x\left[\ {\rm tr}\ G^2+(2\pi\a')^2\ {\rm tr}\ G^4+...\ \right]\ .
\label{5.14}
\eeqn
Rescaling the fields as
\eqn
G\sim \frac{(N_1N_2)^{1/4}g_{YM}}{\a'}\tilde{G}
\sim \frac{R}{\a'}\tilde{G}\ ,\ \ \ 
\tilde{G}=d\tilde{B}+\frac{R}{\a'}[\tilde{B},\tilde{B}]\ ,
\label{5.15}
\eeqn
we obtain the action
\eqn
S\sim-\sqrt{N_1N_2}\frac{1}{\a'^2}
\int d^6x\left[\ {\rm tr}\ \tilde{G}^2+R^2\ {\rm tr}\ \tilde{G}^4+...\ \right]\ .
\label{5.16}
\eeqn
Note that the background field $\tilde{G}^0$ remains finite, i.e.
$\tilde{G}^0\sim {\rm diag}(\tan{\th_1},...)/R$. The $1/\a'^2$ factor in the
front of the action is important because we are compactifying the theory
on $T^4$ with a volume $V_4\sim\a'^2$ and therefore it ensures that
the action remains well defined in the limit $\a'\rightarrow 0$. We are 
keeping $|\tan{\th_i}|\ll 1$ such that our gauge theory fluctuating 
spectrum does not suffer from DBI and derivative corrections.
However, for processes involving energies such that 
$E\sim 1/R$ there will be DBI corrections \cite{Alwis}. In the 
infrared limit $E\ll 1/R$ we recover the SYM description and after
reduction to $1+1$ dimensions we recover the superconformal limit 
in the original derivation of the duality \cite{Mald2}. Also, this
limit corresponds on the supergravity side to the $r\rightarrow 0$ limit 
and we recover the near horizon geometry (moving in $r$ corresponds
to moving in the energy scales on the field theory side \cite{Mald2}).

Our model gives a definite proposal for the conformal theory and
for the coupling of the conformal fields to the bulk fields on the 
$AdS_3$ boundary. In other words Maldacena's duality proposal may
be recasted in the following form: {\em The Higgs branch of the
large $N$ limit of 6-dimensional SYM theory compactified on 
$T^4$ with a 't Hooft twist is dual to supergravity on 
$AdS_3\times S^3\times T^4$}. The parameters
relating the dual theories have already been explained. The coupling to 
the bulk fields is determined by the DBI action.

A more precise formulation of the duality conjecture was given by means 
of calculating conformal field theory correlators using the supergravity
near horizon geometry \cite{Gubs..,Witt3}. Unfortunately the
number of calculations that may be done to test this conjecture is very 
limited because in the
overlapping domain of validity of the dual theories the 't Hooft
coupling of the gauge theory is very large. Also, one would like 
to investigate whether this duality is carried away from the
conformal and near horizon limits. In this case we need the full 
strongly coupled DBI action. Our model is a starting
point to perform such computations in parallel with the D-3-brane
case \cite{Gubs..1,GubsHash} (see also \cite{Teo,Mari} for work on
the D5-D1 brane system). 

In the following we shall argue that in the 't Hooft limit the tension 
of the instanton strings in (\ref{tension}) gets normalised and it 
scales as 
\eqn
T_{eff}\sim \frac{1}{R^2}\sim 
\frac{1}{2\pi \a'g\sqrt{N_1N_2n_L\bar{n}_L}}\equiv
\frac{1}{2\pi \a'g\sqrt{Q_5Q_1}}\ ,
\label{5.17}
\eeqn
confirming the results in \cite{Math,Gubs,HassWadia}. The argument 
is rather heuristic
and by no means rigourous. The compactification of the action (\ref{5.16})
gives a bosonic action of the type
\eqn
S=-\sqrt{N_1N_2} \int d^2x 
\left( G_{rs}(\tilde{\z})
\p_{\s}\tilde{\z}^r\p^{\s}\tilde{\z}^s+\ ...\right)\ ,
\label{5.18}
\eeqn
where ... denote the DBI corrections and the fields $\tilde{\z}$ are 
dimensionless. Now it is hoped that in the limit 
$N_i\rightarrow \infty$ the Feynman rules
that follow from this action should define an effective action
reproducing such rules. Of course this will involve very large Feynman
graphs because the 't Hooft coupling is becoming very large. 
Such effective action should be identified with the rather 
successful effective string action used in the 
computations of scattering amplitudes. In this 
limit the only scale in the problem is $R$.
We are forced to conclude that the effective string tension scales 
as in (\ref{5.17}). Note that we are arguing here that it is the gauge
field $A_{\a}$ that is associated with the effective string 
description. An opposite point of view was advocated in \cite{Alwis}.
One could argue that a tension like
\eqn
T_{eff}\sim
\frac{1}{2\pi \a'g(N_1p_1\bar{p}_1+N_2p_2\bar{p}_2)}\sim
\frac{1}{2\pi \a'gQ_5}\ ,
\label{5.19}
\eeqn
would remain finite in the large $N$ limit. This is in fact true. However,
it is difficult to see how it would arise when considering the Feynman
rules that should originate the effective string action defined above. The
reasons are: Firstly, the fields $b_{\ha}$ associated with the
instanton strings transform under the fundamental representation 
of $U(N_1p_1\bar{p}_1)\otimes U(\overline{N_2p_2\bar{p}_2})$, therefore
the trace of any gauge invariant combination of these fields depends
on $N_i$ through a power of $N_1N_2$ only; Secondly, all couplings in the
action (\ref{5.16}) depend on $N_i$ in the same way. It 
follows that any scattering amplitude is bound to depend on $N_i$ 
through the particular combination $N_1N_2$.

\section{Conclusion}
\news

Let us start by summarising our results. We have argued that a model based
on D-5-branes  with a constant self-dual field strength on $T^4$ describes
5-dimensional black holes within a perturbative string theory framework
and that the D5-D1 system constitutes a special case of this model.
The fluctuating spectrum of this bound state is described by Polchinski's
open strings ending on the D-5-branes. This means that the
Higgs branch of the theory, which describes the ``internal'' excitations
of the bound state, is associated with fundamental modes of the 
worldvolume fields. We may take the volume of 
the $T^4$ to satisfy $V_4\sim \a'$ while string derivative corrections
are negligible. The explicite knowledge of the microscopics of the D-brane
bound state allowed us to make a definite proposal for the conformal 
field theory governing
the Higgs branch of the theory. This means that we may deduce from first
principles the coupling of the bulk fields to the worldvolume fields.
We have done this for a minimally coupled scalar and find agreement with 
the supergravity scattering cross section calculation. Also, the explicite
knowledge of this conformal theory is relevant for Maldacena's 
duality proposal. We think that our model could be a starting point to 
the investigation of the field theory side of this duality conjecture.

Regretfully we have left for the future a number of calculations that 
should prove or disprove the validity of our approach to black hole dynamics.
One should calculate the coupling of
the instanton strings to the minimally coupled scalars arising from the
internal metric on the $T^4$ as well as to the fixed scalars. In these cases
we expect that the fermions will contribute to the corresponding scattering
processes. We may use fermionisation of the bosonic action or find
the supertorons, i.e. the fermionic partners of the toronic excitations
associated with the fields $b_{\a}$ (and $d_m$). An alternative approach
to this problem is to use the string description of the D-brane bound state
and to calculate the corresponding scattering amplitude using string
techniques. This will involve the usual disk diagram with three vertices 
(two of which are on the disk boundary). One should also generalise the 
D-5-brane bound state so that it allows $f=4$ in the D5-D1 brane 
bound state case. It would also be interesting to reproduce this bound
state while suppressing the DBI corrections. Another interesting problem
is to find the hyperk\"{a}hler metric $G_{rs}$ determining the 
superconformal field theory of the instanton strings . This would 
provide us with a better understanding of the interacting
theory. Another problem is to consider a D-5-brane
configuration with instantons and anti-instanton \cite{Dijk..} (in
this case there are tachyonic modes in the spectra which
signal the instability of the configuration). Knowledge of the
interacting theory could be relevant to understand the entropy
formula away from extremality and the dilute gas regime.
One would also like to generalise the field theory description 
with a 't Hooft twist to other compact manifolds, for example $K3$.

Hopefully, a better understanding of this model will shed light on the
string theory approach to black hole physics. 

\section*{Acknowledgements}

I would like to thank Jerome Gauntlett and Bobby Acharya for a very 
informative discussion and Malcolm Perry for discussions and for reading 
the paper. The financial support of FCT (Portugal) under
programme PRAXIS XXI is gratefully acknowledged.

\section*{Note added}

After this work appeared as a pre-print I learned that a similar 
calculation to the one presented in sections 3 and 4 for the 
D-brane emission rates using the DBI action was carried out 
by S.D. Mathur \cite{Math1}.


\newpage

\begin{thebibliography}{40}
\bibitem{Youm}D. Youm, {\em Black holes and solitons in string theory}, 
hep-th/9710046.
\bibitem{Peet}A.W. Peet, {\em The Bekenstein formula and 
string theory (N-brane theory)}, hep-th/9712253.
\bibitem{StroVafa}A. Strominger and C. Vafa, Phys. Lett. {\bf B379}
    (1996) 99.
\bibitem{Mald1}J.M. Maldacena, Nucl. Phys. {\bf B477} (1996) 168.
\bibitem{DasMath}S.R. Das and S.D. Mathur, Nucl. Phys. {\bf B478}
(1996) 561; Nucl. Phys. {\bf B482} (1996) 153. 
\bibitem{MaldStro}J.M. Maldacena and A. Strominger, Phys. Rev. 
{\bf D55} (1997) 861.
\bibitem{Dhar..}A. Dhar, G. Mandal and S.R. Wadia, Phys. Lett. 
{\bf B388} (1996) 51.
\bibitem{KlebGubs1}S.S. Gubser and I.R. Klebanov, Nucl. Phys. {\bf B482}
(1996) 173; Phys. Rev. Lett. {\bf 77} (1996) 4491.
\bibitem{Call..}C.G. Callan, S.S. Gubser, I.R. Klebanov and A.A. Tseytlin,
Nucl. Phys. {\bf B489} (1997) 65.
\bibitem{KlebMath}I.R. Klebanov and S.D. Mathur, Nucl.Phys. {\bf B500}
(1997) 115.
\bibitem{KlebKras}I.R. Klebanov and M. Krasnitz, Phys. Rev. {\bf D55}
(1997) 3250.
\bibitem{MaldStro1}J.M. Maldacena and A. Strominger, Phys. Rev. 
{\bf D56} (1997) 4975.
\bibitem{HawkMari}S.W. Hawking and M.M. Taylor-Robinson, 
Phys. Rev. {\bf D55} (1997) 7680.
\bibitem{Kleb..1}I.R. Klebanov, A. Rajaraman and A.A. Tseytlin,
Nucl.Phys. {\bf B503} (1997) 157.
\bibitem{Math}S.D. Mathur, Nucl. Phys. {\bf B514} (1998) 204.
\bibitem{Gubs}S.S. Gubser, Phys. Rev. {\bf D56} (1997) 4984. 
\bibitem{CvetLars}M. Cveti\v{c} and F. Larsen, Phys. Rev. {\bf D56} 
(1997) 4994; Nucl. Phys. {\bf B506} (1997) 107.
\bibitem{Hoso}K. Hosomichi, Nucl. Phys. {\bf B524} (1998) 312.
\bibitem{Kleb}I.R. Klebanov, Nucl. Phys. {\bf B496} (1997) 231. 
\bibitem{Kleb..}S.S. Gubser, I.R. Klebanov and A.A. Tseytlin,
Nucl. Phys. {\bf B499} (1997) 217. 
\bibitem{KlebGubs}S.S. Gubser and I.R. Klebanov, Phys. Lett.
{\bf B413} (1997) 41.
\bibitem{Polc}J. Polchinski, Phys. Rev. Lett. {\bf 75} (1995) 4724;
{\em TASI Lectures on D-branes}, hep-th/9611050.
\bibitem{CostaPerry}M.S. Costa and M.J. Perry, Nucl. Phys. {\bf B524} 
(1998) 333.
\bibitem{CostaPerry1}M.S. Costa and M.J. Perry, Nucl. Phys. {\bf B520}
(1998) 205.
\bibitem{Mald2}J.M. Maldacena, {\em The Large N limit of superconformal
 field theories and supergravity}, hep-th/9711200.
\bibitem{Gubs..}S.S. Gubser, I.R. Klebanov and A.M. Polyakov, Phys. Lett. 
{\bf B428} (1998) 105. 
\bibitem{Witt3}E. Witten, {\em Anti De Sitter Space And Holography},
hep-th/9802150.
\bibitem{HassWadia}S.F. Hassan and S.R. Wadia, {\em  Gauge theory 
description of D-brane black holes: Emergence of the effective SCFT
and Hawking radiation}, hep-th/9712213.
\bibitem{Witt4}E. Witten, Nucl. Phys. {\bf B460} (1995) 335.
\bibitem{tHoo}G. 't Hooft, Nucl. Phys. {\bf B153}
    (1979) 141; Commun. Math. Phys. {\bf 81} (1981) 267.
\bibitem{Baal}P. Van Baal, Commun. Math. Phys. {\bf 85} (1982) 529;
Commun. Math. Phys. {\bf 94} (1984) 397.
\bibitem{GuraRamg}Z. Guralnik and S. Ramgoolam, Nucl. Phys. {\bf B499} 
(1997) 241; Nucl. Phys. {\bf B521} (1998) 129.
\bibitem{HashTayl}A. Hashimoto and W. Taylor, Nucl. Phys. {\bf B503}
    (1997) 193.
\bibitem{Doug}M. Douglas, {\em Branes within Branes}, hep-th/9512077.
\bibitem{Abou..}A. Abouelsaood, C.G. Callan, C.R. Nappi and S.A. Yost,
    Nucl. Phys. {\bf B280} (1987) 599.
\bibitem{Mald3}J.M. Maldacena, Phys. Rev. {\bf D55} (1997) 7645.
\bibitem{Mald5}J.M. Maldacena, {\em Black Holes in String Theory},
    Ph.D. Thesis, hep-th/9607235.
\bibitem{Dijk..}R. Dijkgraaf, E. Verlinde and H. Verlinde,
Nucl. Phys. {\bf B506} (1997) 121.
\bibitem{Doug..}M. Douglas, J. Polchinski and A. Strominger,
J. High Energy Phys. {\bf 12} (1997) 003.
\bibitem{Kita}Y. Kitazawa, Nucl. Phys. {\bf B289} (1987) 599.
\bibitem{AndrTsey}O.D. Andreev and A.A. Tseytlin, Nucl. Phys. 
{\bf B311} (1988) 205.
\bibitem{CallMald}C. Callan and J.M. Maldacena, Nucl. Phys. {\bf B472}
    (1996) 591.
\bibitem{Ahar}O. Aharony, M. Berkooz, S. Kachru, N. Seiberg and
E. Silverstein, Adv. Theor. Math. Phys. {\bf 1} (1998) 148. 
\bibitem{Witt}E. Witten, J. High Energy Phys. {\bf 07} (1997) 3. 
\bibitem{Witt1}E. Witten, Strings '95 (World Scientific, 1996), ed.
I. Bars et al., 501.
\bibitem{Brec}D. Brecher, {\em BPS States of the Non-Abelian 
Born-Infeld Action}, hep-th/9804180.
\bibitem{Tsey}A.A. Tseytlin, Nucl. Phys. {\bf B501} (1997) 41.
\bibitem{Mald4}J.M. Maldacena, {\em Branes probing black holes}, 
hep-th/9710014.
\bibitem{MaldSuss}J.M. Maldacena and L. Susskind, Nucl. Phys. {\bf B475}
    (1996) 679.
\bibitem{Horo..}G.T. Horowitz, J.M. Maldacena and A. Strominger,
Phys. Lett. {\bf B383} (1996) 151.
\bibitem{Pres..}J. Preskill, P. Schwarz, A. Shapere, S. Trivedi and
F. Wilczek, Mod. Phys. Lett. {\bf A6} (1991) 2353.
\bibitem{HolzWilc}C. Holzhey and F. Wilczek, Nucl. Phys. {\bf B380} 
(1992) 447.
\bibitem{KrausWilc}P.Kraus and F. Wilczek, Nucl. Phys. {\bf B433}
(1995) 403.
\bibitem{Hyun}S. Hyun, {\em U-duality between Three and Higher 
Dimensional Black Holes}, hep-th/9704005.
\bibitem{Boon..}H.J. Boonstra, B. Peeters and K. Skenderis,
Phys. Lett. {\bf B411} (1997) 59.
\bibitem{SfetSken}K. Sfetsos and K. Skenderis, Nucl. Phys. {\bf B517}
(1998) 179.
\bibitem{CousHenn}O. Coussaert and M. Henneaux, Phys. Rev. Lett.
{\bf 72} (1994) 183.
\bibitem{Stro}A. Strominger, J. High Energy Phys. {\bf 02} (1998) 009.
\bibitem{Bana..}M. Banados, C. Teitelboim and J. Zanelli, Phys. Rev.
Lett. {\bf 69} (1992) 1849.
\bibitem{AchuTown}A. Ach\'ucarro and P.K. Townsend, Phys. Lett. 
{\bf B180} (1986) 89.
\bibitem{Witt2}E. Witten, Nucl. Phys. {\bf B311} (1989) 46.
\bibitem{Carl}S. Carlip, Nucl. Phys. {\bf B362} (1991) 111.
\bibitem{BalaLars}V. Balasubramanian and F. Larsen, {\em Near Horizon 
Geometry and Black Holes in Four Dimensions}, hep-th/9802198.
\bibitem{Birm..}D. Birmingham, I. Sachs and S. Sen, Phys. Lett. 
{\bf B424} (1998) 275. 
\bibitem{Lee}T. Lee, {\em Topological Ward identity and anti-de Sitter 
space/CFT correspondence}, hep-th/9805182; {\em The Entropy of the 
BTZ black hole and AdS/CFT correspondence}, hep-th/9806113.
\bibitem{Behr}K. Behrndt, {\em Branes in N=2, D = 4 supergravity and 
the conformal field theory limit}, hep-th/9801058.
\bibitem{Bana..1}M. Banados, T. Brotz and M.E. Ortiz, {\em Boundary dynamics 
and the statistical mechanics of the (2+1)-dimensional black hole},
hep-th/9802076.
\bibitem{Kalo}N. Kaloper, {\em Entropy count for extremal 
three-dimensional black strings},
\newline hep-th/9804062.
\bibitem{MaldStro2}J.M. Maldacena and A. Strominger, {\em AdS(3) black 
holes and a stringy exclusion principle}, hep-th/9804085.
\bibitem{Deger..}S. Deger, A. Kaya, E. Sezgin and P. Sundell, 
{\em Spectrum of D = 6, N=4b supergravity on $AdS_3\times S^3$}, 
hep-th/9804166.
\bibitem{Behr1}K. Behrndt, I. Brunner and I. Gaida, {\em Entropy and 
conformal field theories of AdS(3) models}, hep-th/9804159; 
{\em AdS(3) gravity and conformal field theories}, hep-th/9806195.
\bibitem{CvetLars1}M. Cveti\v{c} and F. Larsen, {\em Near Horizon 
Geometry of Rotating Black Holes in Five Dimensions}, hep-th/9805097;
{\em Microstates of Four-Dimensional Rotating Black Holes from 
Near-Horizon Geometry}, hep-th/9805146.
\bibitem{Bana..2}M. Banados, K. Bautier, O. Coussaert, M. Henneaux 
M. Ortiz, {\em Anti-de Sitter/CFT correspondence in three-dimensional 
supergravity}, hep-th/9805165.
\bibitem{Evans..}J.M. Evans, M.R. Gaberdiel and M.J. Perry,
{\em The no ghost theorem for AdS(3) and the stringy exclusion principle},
hep-th/9806024.
\bibitem{Bana..3}M. Banados and M.E. Ortiz, {\em The Central charge 
in three-dimensional anti-de Sitter space}, hep-th/9806089.
\bibitem{Lars}F. Larsen, {\em The Perturbation Spectrum of 
Black Holes in N=8 Supergravity}, hep-th/9805208; {\em Anti-DeSitter 
Spaces and Nonextreme Black Holes}, hep-th/9806071.
\bibitem{Boer}J. de Boer, {\em Six-dimensional supergravity on 
$S^3\times AdS_3$ and 2-D conformal field theory}, hep-th/9806104.
\bibitem{Hwang}S. Hwang. {\em Unitarity of strings and noncompact 
Hermitian symmetric spaces}, hep-th/9806049.
\bibitem{Empa..}R. Emparan and I. Sachs, {\em Quantization of 
AdS(3) black holes in external fields}, hep-th/9806122.
\bibitem{Give..}A. Giveon, D. Kutasov and N. Seiberg, 
{\em Comments on String Theory on $AdS_3$}, hep-th/9806194.
\bibitem{Alwis}S.P. Alwis, {\em Supergravity the DBI Action and Black 
Hole Physics}, hep-th/9804019.
\bibitem{Gubs..1}S.S. Gubser, A. Hashimoto, I.R. Klebanov and 
M. Krasnitz, {\em Scalar absorption and the breaking of the world 
volume conformal invariance}, hep-th/9803023.
\bibitem{GubsHash}S.S. Gubser and A. Hashimoto, {\em Exact absorption 
probabilities for the D3-brane}, hep-th/9805140.
\bibitem{Teo}E. Teo, {\em Black hole absorption cross-sections and the 
anti-de Sitter-conformal field theory correspondence}, hep-th/9805014.
\bibitem{Mari}M.M. Taylor-Robinson, {\em The D1-D5 brane system in 
six dimensions}, hep-th/9806132.
\bibitem{Math1}S.D. Mathur, private communication.
\end{thebibliography}
\end{document}